\documentclass[journal]{rmaa}
\usepackage{aas_macros}

\usepackage{paralist}

\usepackage{psfrag,color}

\usepackage[latin1]{inputenc}
\usepackage[T1]{fontenc}
\usepackage{rotating}

\title{Long term CCD photometry of the distant cluster NGC2419: the CMD revisited} 

\author{
  A. Arellano Ferro,\altaffilmark{1}
  S, Muneer,\altaffilmark{2}
     Sunetra Giridhar,\altaffilmark{2}
 I. Bustos Fierro,\altaffilmark{3}
    M. A. Yepez,\altaffilmark{1,4}
G.A. Garc\'ia P\'erez,\altaffilmark{5}
G. Rios Segura.\altaffilmark{5}
}

\altaffiltext{1}{Instituto de Astronom\'ia, Universidad Nacional Aut\'onoma de M\'exico, Ciudad Universitaria, C.P. 04510, M\'exico.}

\altaffiltext{2}{Indian Institute of Astrophysics, Bangalore, India.}

\altaffiltext{3}{Observatorio Astron\'omico, Universidad Nacional de C\'ordoba, C\'ordoba C.P. 5000, Argentina.}

\altaffiltext{4}{Instituto Nacional de Astrof\'isica, \'Optica y Electr\'onica (INAOE), Luis Enrique Erro No.1, Tonantzintla, Pue., C.P. 72840, M\'exico}

\altaffiltext{5}{Facultad de F\'isica, Universidad Veracruzana, Xalapa, M\'exico.}
\shortauthor{A. Arellano Ferro ET AL.}
\shorttitle{NGC 2419: the CMD revisited}

\abstract{Employing \emph{VI} images of NGC 2419 acquired over 17 years, light curves for most of the known variables in the field of the cluster are produced.  A cluster membership analysis for about 3100 stars in the cluster field with proper motions from $Gaia$-DR3, revealed the presence of member stars as far as 140 pc from the cluster center and enabled the construction of a cleaner CMD free of field stars. 
It was found that RRab and RRc stars share the inter-order region in the instability strip, which is unusual for OoII clusters.
Theoretical considerations confirm that Pop II cepheids are descendants of extreme ZAHB blue tail stars with very thin envelopes of about 10\% of the total mass.
Member RR Lyrae stars were employed to calculate independent estimates of the mean cluster metallicity and distance; we found [Fe/H]$_{\rm UV}= -1.90 \pm 0.27$ and $D=86.3 \pm 5.0$ kpc from the RRab 
and [Fe/H]$_{\rm UV}= -1.88 \pm 0.30$ and $D=83.1 \pm 8.1$ kpc from the RRc light curves.}

\resumen{Empleando im\'agenes CCD \emph{VI} del c\'umulo globular NGC 2419, obtenidas durante 17 a\~nos, se hemos  construido curvas de luz de gran parte de las estrellas variables en el campo del c\'umulo. A partir de un an\'alisis de membres\'ia basado en los movimientos propios de $Gaia$-DR3 de 3100 estrellas, detectamos miembros a distancias de 140 pc del centro del c\'umulo, y construimos un Diagrama Color-Magnitud libre de estrellas de campo. Encontramos que los dos modos de pulsaci\'on RRab y RRc comparten la regi\'on bimodal de la zona de inestabilidad, lo cual es inesperado en c\'umulos Oo II. Nuestros modelos confirman que las estrellas cefeidas de Pob II provienen del extremo azul de la ZAHB con envolventes muy delgadas, de $\sim$ 10\% de la masa total.
A partir de estrellas RR Lyrae miembros calculamos la metalicidad y distancia medias del c\'umulo
[Fe/H]$_{\rm UV}= -1.90 \pm 0.27$ y $D=86.3 \pm 5.0$ kpc para estrellas RRab,
y [Fe/H]$_{\rm UV}= -1.88 \pm 0.30$ y $D=83.1 \pm 8.1$ kpc para estrellas RRc.}

\addkeyword{Globular clusters: general} 

\addkeyword{Globular clusters: individual (NGC 2419)} 

\addkeyword{Stars: horizontal branch}
\addkeyword{Stars: distances}

\addkeyword{Stars: fundamental parameters}

\addkeyword{Stars: variables: RR Lyrae}

\begin{document}
\maketitle

\section{Introduction}
\label{sec:Intro}

NGC 2419  (C0734+390 in the IAU nomenclature) ($\alpha = 07^{\mbox{\scriptsize h}}
38^{\mbox{\scriptsize m}} 08.47^{\mbox{\scriptsize s}}$, $\delta = +38^{\circ} 52\arcmin
56.8\arcsec$, J2000; $l = 180.37^{\circ}$, $b =+25.24^{\circ}$) is a large and very luminous globular cluster at  about 80 kpc  from the Galactic center and hence it is among the most distant clusters in the outer halo of our Galaxy. It is noted for being a very loose or extended system given its brightness $M_V$ and half-light radius, which led \citet{vdB2004} to suggest its extra galactic origin, and the likelihood of it being a stripped core of a former spheroidal dwarf galaxy. These suggestion however has not been supported by \citet{Ripepi2007} on the ground of the following arguments; the cluster is of the Oo II type while extra galactic clusters and dwarf galaxies reside within the Oosterhoff gap \citep{Catelan2009}, the lack of substructures in the main sequence bellow the TO (i.e. lack of multiple populations), the thinness of the red giant branch (ruling out multiple chemical compositions among cluster stars), and the apparent absence of extra-tidal structures. In Ripepi et al's opinion NGC 2419 is a normal metal-poor Galactic globular cluster.

The variable stars population of NGC 2419 is quite rich. There are 101 variables registered in the Catalogue of Variable Stars in Globular Clusters (CVSGC) \citep{Clement2001}, 75 of which are RR Lyrae, 1 Pop II or CW star, 12 SX Phe, 5 long-period red giants, 3 eclipsing binaries, 2 $\delta$ Scuti stars and  3 non-classified.
A thorough analysis of these variable stars, based on high quality images from the HST, and two large ground telescopes; the Galileo (TNG) 3.5 and the Subaru 8.2m was carried out by \citet{diCriscienzo2011}.

\begin{table}[t]
\scriptsize
\begin{center}
\caption{Log of observations of NGC 2419.}
\label{log}
    \begin{tabular}{cccccc} 
    \hline
    Date & N$_V$ & t$_V$(s) & N$_I$ & t$_I$(s) & Mean seeing ($^{\prime \prime}$) \\
    \hline
    01/04/2005 & 7   & 600 & - & - & 1.9 \\
    02/04/2005 & 8   & 600 & - & - & 1.7 \\
    03/04/2005 & 11  & 600 & - & - & 1.9 \\
    19/01/2006 & 14  & 600 & - & - & 3.4 \\
    09/03/2007 & 8   & 600 & - & - & 2.0 \\
    10/03/2007 & 18  & 600 & - & - & 2.7 \\
    10/04/2007 & 10  & 600 & - & - & 2.0 \\
    11/04/2007 & 13  & 600 & - & - & 2.0 \\
    07/01/2009 & 5   & 300-600 & 2 & 300     & 1.7\\
    08/01/2009 & 2   & 600     & 2 & 300     & 2.3 \\
    15/12/2011 & 3   & 450-500 & 4  & 100-150 & 3.5 \\
    16/12/2011 & 7   & 400-450 & 7  & 120-150 & 2.8 \\
    17/12/2011 & 2   & 380-420 & 4  & 110-120 & 2.7 \\
    05/02/2012 & 22  & 180-300 & 20 & 90      & 2.2 \\
    28/02/2012 & 32  & 180-200 & 35 & 80-100  & 2.1 \\
    29/02/2012 & 16  & 250-450 & 17 & 80-250  & 2.8 \\
    01/03/2012 & 11  & 180-300 & 12 & 80-150  & 2.7 \\
    19/01/2013 & 25  & 150-400 & 27 & 75-150  & 2.1 \\
    20/01/2013 & 41  & 170-210 & 44 & 80-100  & 2.2 \\
    21/01/2013 & 3   & 180     & 5  & 80      & 2.4 \\
    04/03/2013 & 6   & 125-150 & 6  & 65-85   & 1.8 \\
    01/02/2017 & 14  & 180     & 13 & 80      & 2.5 \\
    02/02/2017 & 14  & 180     & 13 & 80      & 2.7 \\
    16/02/2020 & 16  & 200     & 16 & 100     & 2.8 \\
    17/02/2020 & 8   & 200     & 8  & 100     & 2.4 \\
    19/03/2021 & 12  & 200     & 12 & 100     & 1.9 \\
    04/11/2021 & 27  & 200     & 28 & 100     & 2.6 \\
    12/02/2021 & 10  & 200     & 10 & 100     & 2.0 \\
    03/01/2022 & 8   & 200     & 8  & 100     & 2.9 \\
    31/01/2022 & 32  & 200     & 34 & 100     & 2.4 \\
    \hline
    Total: & 405 & & 327 & & \\
    \hline
\end{tabular}
Columns N$_V$ and N$_I$ give the number of images taken with the $V$ and $I$ filters respectively. Columns t$_V$ and t$_I$ provide the exposure time, or range of exposure times. In the last column the average seeing is listed.
\end{center}
\end{table}

Our team has obtained \emph{VI} CCD images of NGC 2419 between 2005 and 2022 for a total of 405 and 327 in $V$ and $I$ respectively, using the 2m telescope of Indian Astronomical Observatory (IAO). Naturally the depth and accuracy of our photometry is not comparable to that used by \citet{diCriscienzo2011}, and due to crowdiness we were not able to recover the light curves for all known variables. It may seem pretentious to draw conclusions from RR Lyrae stars at $V\sim 20.5$, which nears the faint limit of our data. Nevertheless, we can use our photometry to provide a new discussion of the cluster membership of the stars in the field of our images, particularly those of variable nature, and provide independent estimates from the mean cluster reddening, average metallicity and distance. 

Our approach to the determination of mean $M_V$ and [Fe/H] of RRL stars, is the Fourier light curve decomposition, of both the fundamental mode and first overtone pulsators RRab and RRc respectively. This, and the employment of well established semi-empirical calibrations and their zero points between the Fourier parameters and the physical quantities, provide individual stellar estimations of the distance and [Fe/H], hence the average of these values for tested cluster member variables lead to a proper average values for parent cluster. The present paper is a report of our results.

\begin{figure}[b!]
\begin{center}
\includegraphics[width=7.9cm]{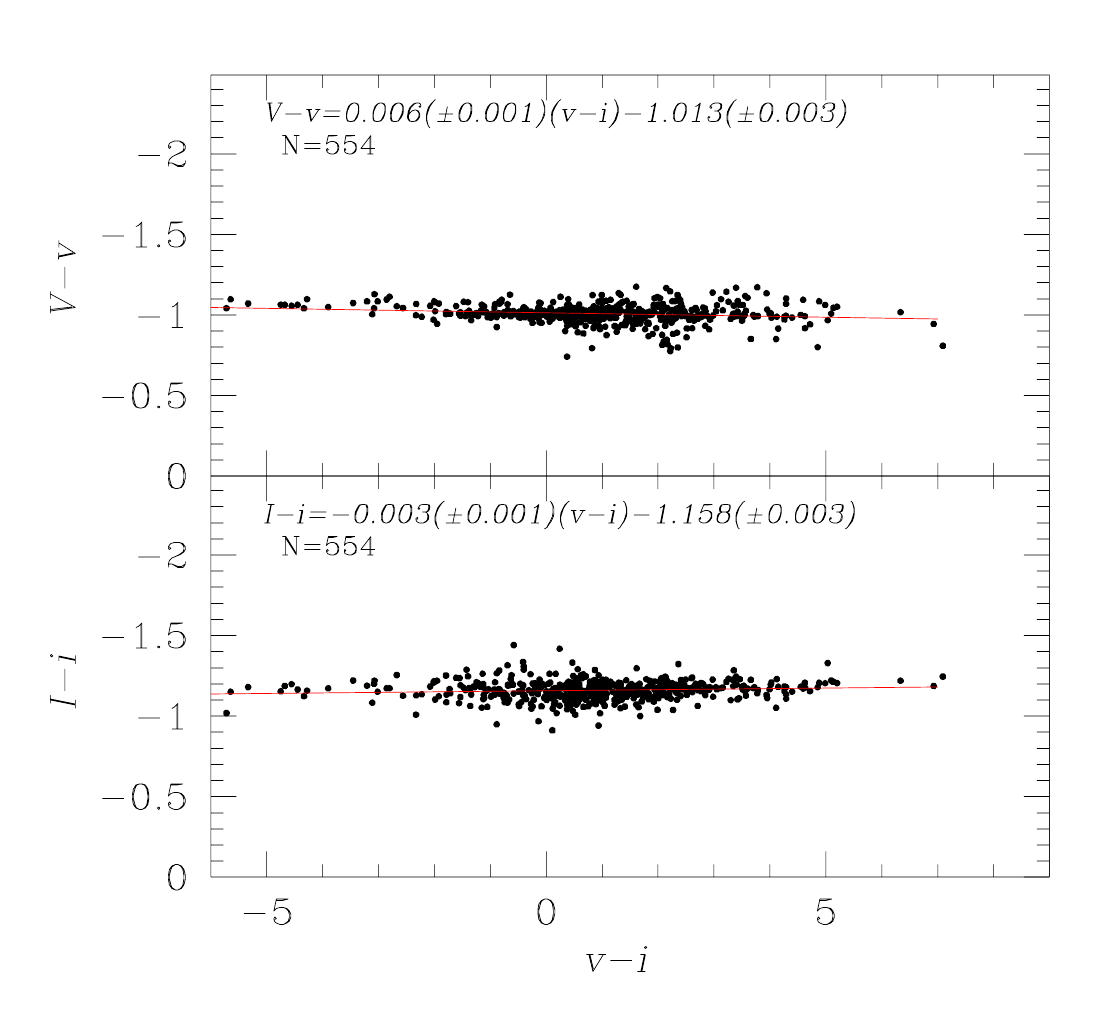}
\caption{Transformation relationship between instrumental and standard photometric systems. These equations were calculated using 554 local standard stars from the collection of \citet{Stetson2000}.}
\label{Trans_color}
\end{center}
\end{figure}

\section{Observations and image reductions}
\label{sec:ObserRed}

\begin{figure*}[ht]
\begin{center}
\includegraphics[width=16.0cm]{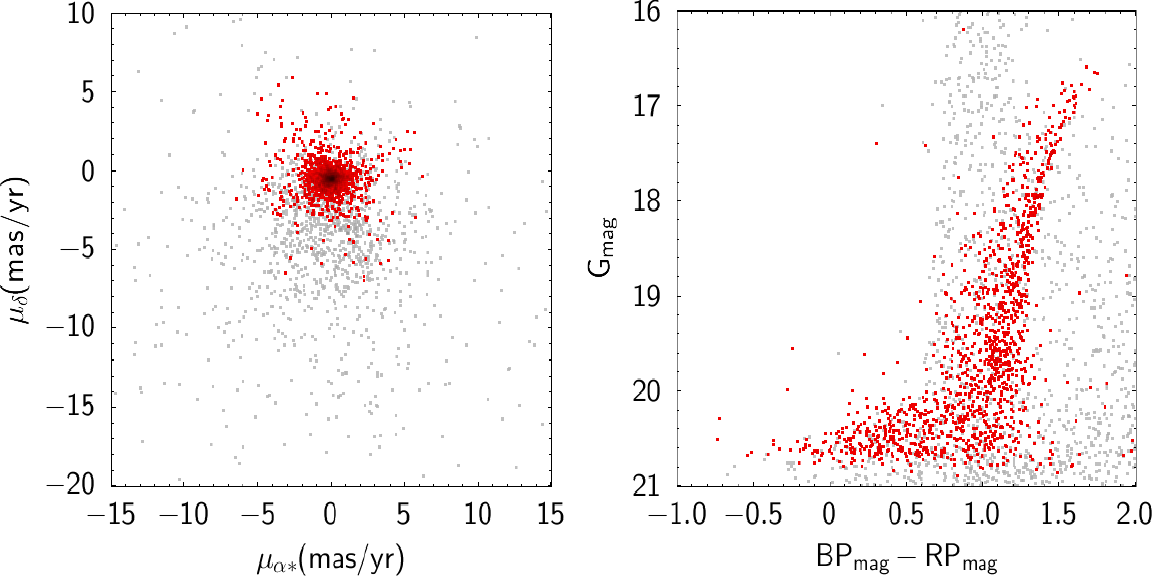}
\caption{$Gaia$-DR3 VPD (left panel) and CMD (right panel) of the cluster NGC 2419. Red and gray points correspond to likely cluster members and field stars respectively, determined as described in section \ref {sec:membership}. A total of 3965 $Gaia$ point sources within 15 arc minutes are displayed, while 1584 were found to be cluster members. }
\label{VPD}
\end{center}
\end{figure*}

\subsection{Observations}

All observations were carried out with the Himalayan Chandra 2.0m Telescope of the Indian Astrophysical Observatory (IAO) at Hanle, in the Indian Himalayan range at about 4500 m above sea level. The detailed log of the observations is given in Table \ref{log} where exposure times and estimations of the prevailing nightly average seeing are indicated. A total of 405 and 327 images in $V$ and $I$ were secured in a time span of seventeen years.

\subsection{Transformation to the Standard System}

\label{Tranformation}

The instrumental photometry was transformed  to the standard  
 Johnson-Kron-Cousins photometric system \citep{Landolt1992}
\emph{VI}, using local standard stars in the fields of the target clusters. These standard stars have been taken from the extensive collection of
\citet{Stetson2000}\footnote{%
 \texttt{https://www.canfar.net/storage/list/STETSON/Standards}}. We found 554 standard stars with instrumental light curves in the field of our images. The standard minus instrumental magnitudes and the light dependence with the $(v-i)$ colour is displayed in Fig. \ref{Trans_color}. The corresponding transformation equations are also given in the figure.

\subsection{Difference Image Analysis}
\label{DIA}

All the image photometric treatment has been performed using the Difference Image Analysis using the \emph{DanDIA} pipeline \citep{Bramich2008,Bramich2013,Bramich2015}. The approach and its caveats have been described in detailed by \citet{Bramich2011}.

\section{Stellar membership analysis}
\label{sec:membership}

Distinguishing the true cluster members from the field stars projected on the field of view of a cluster is relevant since one is interested in a clean CMD that represents the structure of the system. Presently, this challenge is on reach given the high quality proper motions available in the $Gaia$ mission \citep{Gaia2023}.

The method developed by \citet{Bustos2019} to determine the stellar membership is based on a two step approach: 1) it finds groups of stars with similar characteristics in the four-dimensional space of the gnomonic coordinates ($X_{\rm t}$,$Y_{\rm t}$) and proper motions ($\mu_{\alpha*}$,$\mu_\delta$) employing the BIRCH clustering algorithm \citep{Zhang1996} and 2) in order to extract likely members that were missed in the first stage, the analysis of the projected distribution of stars with different proper motions around the mean proper motion of the cluster is performed.

\begin{figure*}
\begin{center}
\includegraphics[width=16.0cm]{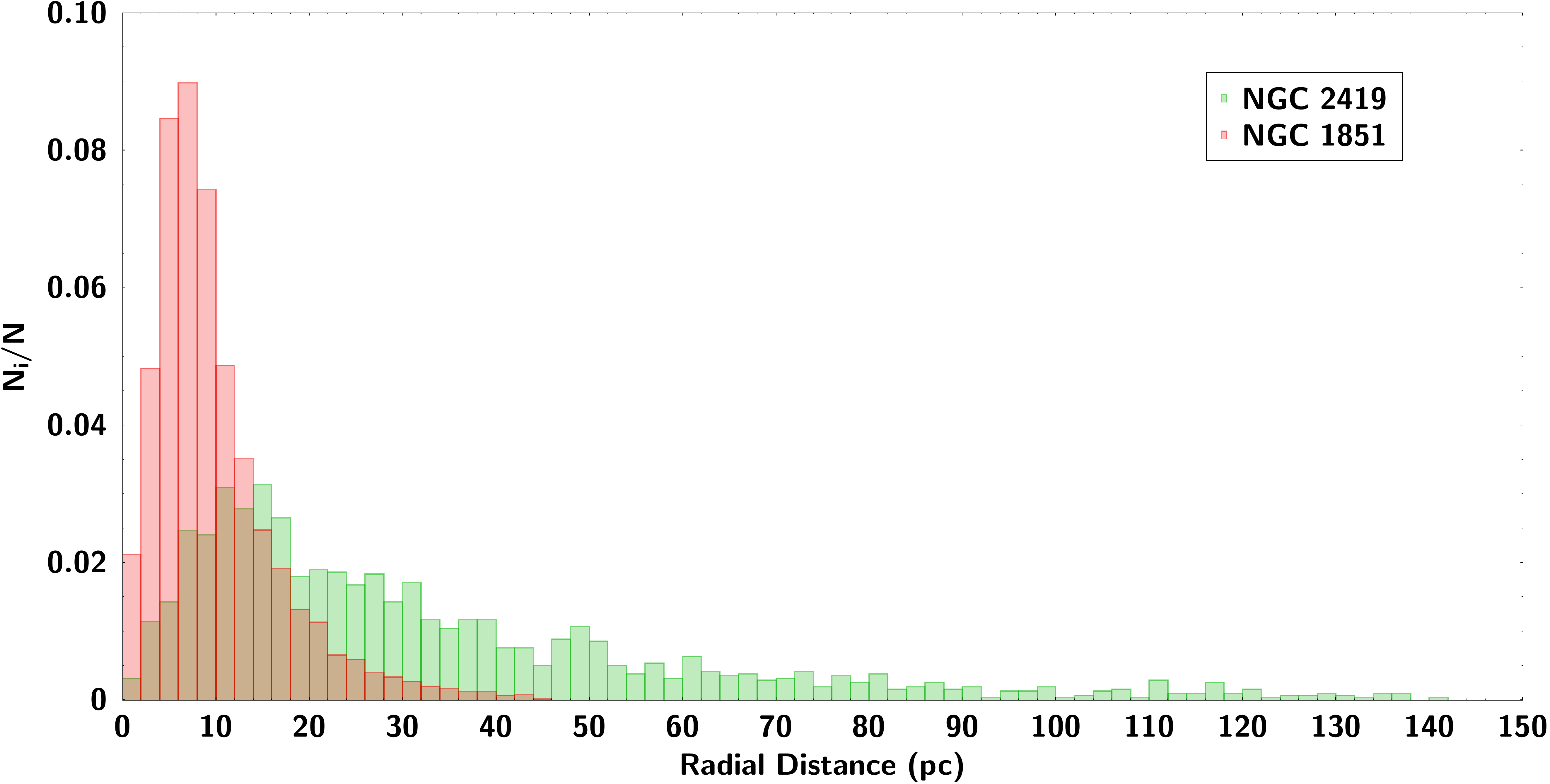}
\caption{Star members distribution of the extended cluster NGC 2419 compared with the large halo cluster NGC 1851.  The horizontal axis represents the radial distance projected on the plane of the sky.}
\label{histogram}
\end{center}
\end{figure*}

This method was applied to the stars within a radius of 15 arc minutes from the center of NGC 2419. This field contains 3965 $Gaia$ sources but only 3129 have a measurement of the proper motion, 1584 of which were found to be likely cluster members. In Fig. \ref{VPD} the corresponding Vector Point Diagram (VPD) and colour-magnitude diagram (CMD) showing the cluster member and field stars are shown.
The farthest member from the center is about 6 arc minutes away which, at a distance of 84.0 kpc (see our distance determination in section \ref{sec:Four}) correspond to a distance of about 140 pc.  In spite of its half-light radius at about 17.9 pc, comparable to other clusters in the halo \citep{vdB2004}, these star members at such large distances are the indication of the extended halo that led \citet{vdB2004} to suggest its extra galactic origin and presently being the remains of a former dwarf galaxy tidally stripped by the Milky Way. In Fig. \ref{histogram}, the radial distance distribution of member stars in NGC 2419 is displayed and shows its extended nature. For comparison we included the similar distribution of large halo cluster NGC 1851.
The membership analysis reveals the bound nature of a subtle cluster halo as large as 140 pc from the cluster center.

We have been able to obtain \emph{VI} photometry for 1107 point sources in the field of our images. These light curves include 74 of the known variables in the cluster \citep{Clement2001}, and the general data for these variables is included in Table \ref{tab:datosgenerales1}. A CMD with these variables identified has been produced and will be discussed in the following sections.

\begin{table*}
    \begin{center}
    \caption{General data of the variables of the FoV in NGC 2419.}
    \footnotesize
    \begin{tabular}{ccccccccccc}
    \hline
    Variable & Type & $<$\textit{V}$>$ & $<$\textit{I}$>$ & $A_{V}$ & $A_{I}$ & Period & $HJD_{max}$ & RA & Dec. & Membership \\
     &   & (mag) & (mag) & (mag) & (mag) & (days) & (d+2450000) & (J2000.0) & (J2000.0) & (m/f/un)\\
    \hline
    &&&&&RRab&&&&& \\
\hline
    V2 & RRab & 19.675 & 18.927 & 0.772 & - &0.792335  & 4169.173 & 07:38:07.96 & +38:52:33.7 & m \\
    V3 & RRab & 20.370 & 19.811 & 1.163 & 0.713 & 0.625995 & 9293.321 & 07:38:12.71 & +38:52:27.4 & m \\
    V5 & RRab & 19.895 & 19.171 & 0.520 & 0.342 & 0.655855 & 9293.166 & 07:38:11.14 & +38:53:38.8 & m \\
    V7 & RRab & 20.484 & 19.886 & 1.341 & 0.830 & 0.627353 & 6312.483 & 07:38:16.19 & +38:54:18.8 & m \\
    V9 & RRab & 20.316 & 19.614 & 1.239 & 0.751 & 0.644727 & 4202.120 & 07:38:05.63 & +38:54:21.2 & m \\
    V11 & RRab & 20.436 & 20.031 & 1.097 & 0.931 & 0.589179 & 4170.333 & 07:38:16.43 & +38:52:42.3 & m \\
    V12 & RRab & 20.405 & 19.871 & 1.087 & 0.687 & 0.661853 & 9611.116 & 07:38:19.79 & +38:54:41.0 & m \\
    V13 & RRab & 20.425 & 19.849 & 1.051 & 0.687 & 0.640 & 4169.334 & 07:38:16.93 & +38:52:40.2 & m \\
    V14 & RRab & 20.257 & 19.777 & 1.310 & 1.003 & 0.741 & 9258.370 & 07:37:58.39 & +38:52:42.3 & m \\
    V15 & RRab & 20.438 & 19.846 & 1.265 & 0.813 & 0.640 & 9293.166 & 07:38:13.66 & +38:53:30.6 & m \\
    V16 & RRab & 20.202 & - & 1.139 & - & 0.666 & 4202.120 & 07:38:12.38 & +38:54:03.4 & m \\
    V17 & RRab & 20.406 & 19.822 & 0.979 & 0.619 & 0.649 & 4201.162 & 07:38:17.74 & +38:54:41.0 & m \\
    V19 & RRab & 20.434 & 19.890 & 1.303 & 0.866 & 0.703 & 9258.370 & 07:37:59.02 & +38:52:15.0 & m \\
    V21 & RRab & 20.318 & 19.830 & 1.497 & 0.690 & 0.686 & 9293.166 & 07:38:03.58 & +38:53:23.7 & m \\
    V22 & RRab & 20.404 & - & 1.123 & - & 0.577 & 6312.478 & 07:38:17.65 & +38:52:44.4 & m \\
    V23 & RRab & 19.976 & 19.476 & 0.996 & - & 0.626 & 5988.205 & 07:38:10.70 & +38:54:10.7 & m \\
    V24 & RRab & 20.427 & 19.743 & 0.869 & 0.799 & 0.653 & 4169.276 & 07:37:55.61 & +38:52:45.7 & m \\
    V25 & RRab & 20.330 & 19.595 & 0.851 & 0.536 & 0.636 & 3464.280 & 07:38:03.28 & +38:53:31.7 & m \\
    V26 & RRab & 19.609 & 18.958 & 0.580 & 0.373 & 0.664 & 4201.163 & 07:38:02.20 & +38:52:03.2 & m \\
    V29 & RRab & 20.349 & 19.909 & 0.759 & 0.598 & 0.726 & 3464.173 & 07:38:03.26 & +38:52:46.9 & m \\
    V30 & RRab & 20.600 & - & 1.152 & - & 0.584 & 4170.189 & 07:38:06.09 & +38:53:16.2 & m \\
    V32 & RRab & 20.188 & 19.443 & 0.729 & 0.615 & 0.642 & 4170.230 & 07:38:06.70 & +38:53:40.8 & m \\
    V35 & RRab & 20.797 & - & 1.627 & - & 0.677 & 4839.390 & 07:38:12.01 & +38:52:59.9 & m \\
    V36 & RRab & 20.310 & - & 0.708 & - & 0.648 & 9611.423 & 07:38:10.30 & +38:53:35.3 & m \\
    V37 & RRab & 19.361 & 18.467 & 0.184 & 0.226 & 0.661 & 4170.257 & 07:38:11.20 & +38:53:09.1 & un \\
    V40 & RRab & 18.930 & 17.989 & 0.229 & 0.195 & 0.576 & 6312.400 & 07:38:13.09 & +38:52:47.2 & un \\
    V42 & RRab & 19.787 & - & 0.277 & - & 0.775 & 5963.194 & 07:38:08.72 & +38:52:45.7 & un \\
    V57 & RRab & 20.041 & 19.516 & 0.812 & 0.483 & 0.736 & 5963.078 & 07:38:06.00 & +38:52:42.7 & m \\
    V59 & RRab & 20.258 & - & 0.585 & - & 0.829 & 5963.161 & 07:38:10.43 & +38:53:09.5 & m \\
    V64 & RRab & 20.385 & 19.638 & 0.337 & 0.449 & 0.781 & 4170.333 & 07:38:10.29 & +38:52:16.3 & m \\
    \hline
    &&&&&RRc&&&&& \\
\hline
    V4 & RRc & 20.420 & 19.917 & 0.403 & 0.380 & 0.392 & 6313.385 & 07:38:15.13 & +38:52:35.1 & m \\
    V6 & RRc & 20.349 & - & 0.553 & - & 0.372 & 3463.242 & 07:38:12.89 & +38:50:43.8 & m \\
    V27 & RRc & 20.454 & - & 0.311 & - & 0.342 & 9258.425 & 07:38:09.78 & +38:51:09.0 & m \\
    V31 & RRc & 20.297 & - & 0.509 & - & 0.388 & 9293.285 & 07:38:21.24 & +38:50:22.9 & m \\
    V33 & RRc & 20.604 & - & 0.427 & - & 0.303 & 5963.118 & 07:38:12.27 & +38:52:33.9 & m \\
    V34 & RRc & 20.437 & - & 0.531 & - & 0.399 & 4170.119 & 07:38:10.34 & +38:55:29.1 & m \\
    V38 & RRc & 18.748 & 17.819 & 0.134 & 0.162 & 0.364 & 5963.161 & 07:38:08.13 & +38:52:03.9 & un \\
    V41 & RRc & 19.324 & 18.629 & 0.167 & 0.154 & 0.396 & 4202.187 & 07:38:07.20 & +38:53:23.1 & un \\
    V48 & RRc & 19.452 & 18.241 & 0.218 & 0.219 & 0.375 & 6356.102 & 07:38:10.04 & +38:52:41.2 & un \\
    V51 & RRc & 19.914 & 19.307 & 0.266 & 0.417 & 0.348 & 4839.375 & 07:38:06.19 & +38:52:57.5 & un \\
    V55 & RRc & 20.392 & - & 0.409 & - & 0.378 & 4839.375 & 07:38:05.99 & +38:52:59.6 & m \\
    V56 & RRc & 19.415 & 19.154 & 0.108 & 0.373 & 0.333 & 5963.117 & 07:38:07.99 & +38:52:24.7 & un \\
    V60 & RRc & 20.376 & - & 0.362 & - & 0.390 & 6313.478 & 07:38:06.66 & +38:53:18.8 & m \\
    V66 & RRc & 20.515 & - & 0.419 & - & 0.387 & 6313.156 & 07:38:11.60 & +38:53:10.5 & m \\
    V67 & RRc & 19.978 & - & 0.305 & - & 0.348 & 5963.118 & 07:38:10.97 & +38:53:23.5 & m \\
    V68 & RRc & 20.333 & - & 0.435 & - & 0.365 & 6313.454 & 07:38:12.15 & +38:52:49.0 & m \\
    V69 & RRc & 20.109 & 19.441 & 0..335 & 0.386 & 0.344 & 5986.225 & 07:38:10.64 & +38:53:37.7 & un \\
    V72 & RRc & 20.412 & 19.700 & 0.426 & 0.350 & 0.415 & 5912.464 & 07:38:09.92 & +38:53:49.5 & m \\
    \hline
    \hline
    \end{tabular}
    \label{tab:datosgenerales1}
     Columns 3 and 4 are means magnitudes, columns 5 and 6 are light curves amplitudes. Column 11 indicates the cluster membership status; m - members, f - field, un - unknown.
    \end{center}
\end{table*}

\begin{table*}
\addtocounter{table}{-1}
\caption{Continue}
\footnotesize
    \begin{center}
   \begin{tabular}{ccccccccccc}
    \hline
    Variable & Type & $<$\textit{V}$>$ & $<$\textit{I}$>$ & $A_{V}$ & $A_{I}$ & Period & $HJD_{max}$ & RA & Dec. & Memb \\
     &   & (mag) & (mag) & (mag) & (mag) & (days) & (d+2450000) & (J2000.0) & (J2000.0) & (m/f/un)\\
    \hline
    V74 & RRc & 20.598 & - & 0.206 & - & 0.309 & 5963.118 & 07:38:13.78 & +38:52:47.9 & m \\
    V75 & RRc & 20.288 & - & 0.429 & - & 0.324 & 4840.363 & 07:38:03.54 & +38:52:16.9 & m \\
    V76 & RRc & 20.128 & 19.579 & 0.275 & 0.302 & 0.324 & 5986.267 & 07:38:12.11 & +38:51:59.2 & m \\
    V77 & RRc & 20.298 & - & 0.332 & - & 0.381 & 5963.117 & 07:38:13.14 & +38:52:08.2 & m \\
    V82 & RRc & 19.911 & 19.594 & 0.205 & 0.327 & 0.343 & 9293.289 & 07:38:13.85 & +38:53:37.3 & m \\
    V84 & RRc & 20.046 & 19.165 & 0.287 & 0.301 & 0.329 & 5986.225 & 07:38:01.46 & +38:53:12.0 & m \\
    V89 & RRc & 20.273 & - & 0.283 & - & 0.287 & 5963.118 & 07:38:19.78 & +38:55:07.4 & f \\
    V90 & RRc & 20.354 & - & 0.328 & - & 0.391 & 5963.124 & 07:38:23.14 & +38:54:11.6 & m \\

\hline
    &&&&&RRd&&&&& \\
\hline
    V39 & RRd & 20.335 & - & - & - & 0.814 & 5963.181 & 07:38:10.72 & +38:50:51.1 & m \\
\hline
    &&&&&RGBs&&&&& \\
\hline
    V1 & RGB & 17.188 & 15.664 & 0.451 & 0.472 & 193.850 & 5963.161 & 07:38:11.61 & +38:51:59.0 & m \\
    V8 & RGB & 17.301 & 15.887 & 0.603 & 0.613 & 16.350 & 9258.370 & 07:38:06.84 & +38:53:34.1 & m \\
    V10 & RGB & 17.085 & 15.552 & 0.466 & 0.374 & 20.800 & 7786.166 & 07:38:09.89 &+38:52:01.1 & m \\
    V20 & RGB & 17.425 & 16.680 & 0.427 & 0.528 & 48.980 & 5963.161 & 07:38:05.96 & +38:53:38.2 & m \\
    V86 & RGB & 17.343 & 15.932 & 0.548 & 0.418 & 49.860 & 5963.161 & 07:38:19.62 & +38:53:15.7 & m \\
  
\hline
    &&&&&V Vir&&&&& \\
\hline    
    V18 & W VIR & 18.837 & 18.207 & 0.412 & 0.725 & 1.579 & 4202.120 & 07:38:07.17 & +38:54:46.8 & m \\

    \hline
    \end{tabular}
    \end{center}
\end{table*}

\section{Variable stars measured in present study}
\label{Vars}

Given the size of our telescope and sky conditions at the time of the observations that limited the deepness and resolution of our imaging data, hence some of the well known variables remain blended and it was not 
possible to measure them well. It should be noted that the HB level of this distant cluster is fainter than 20th magnitude, which is close to the faint limit of our capabilities. Variables in Table \ref{tab:datosgenerales1} are mostly  RR Lyrae (RRab and RRc), but also a sample of Red Giant Branch variables  (RGBs) is available, plus one W Vir (V18) and one double mode or RRd star (V39). The identifications of these variables are given in the charts of Fig. \ref{IdChart}. Their light curves are displayed in the Appendix, where we have distinguished the data from the several observing runs. 

\begin{figure*}[ht]
\begin{center}
\includegraphics[width=12cm]{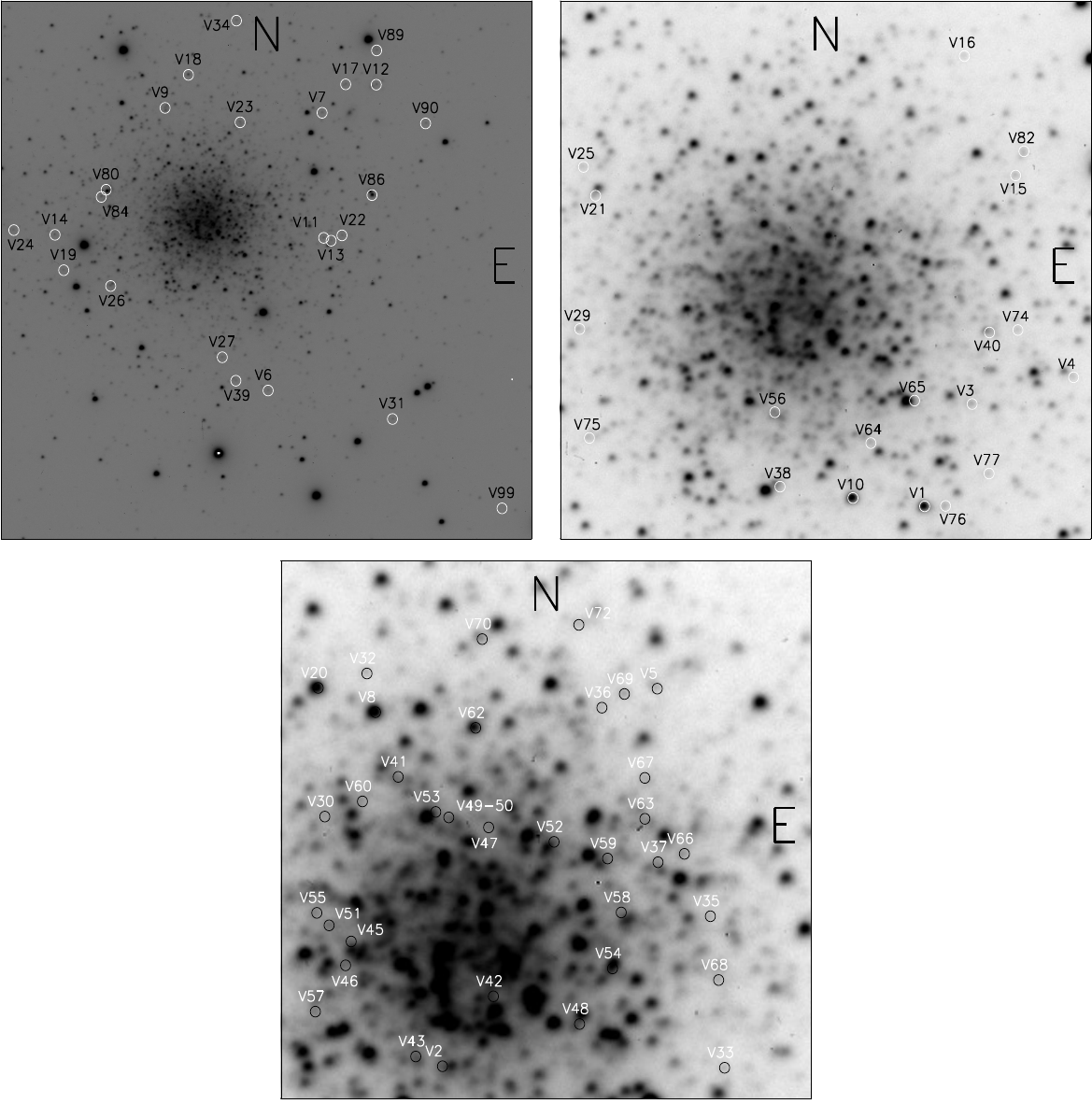}
\caption{Identification chart of variable stars in the field of NGC 2419. The approximate size of each images is $6.9\times6.9$, $2.5\times2.5$ and $1.6\times1.6$ square arc minutes.}
\label{IdChart}
\end{center}
\end{figure*}

\section{The Oosterhoff type of NGC 2419}

\begin{figure}
\begin{center}
\includegraphics[width=6.0cm]{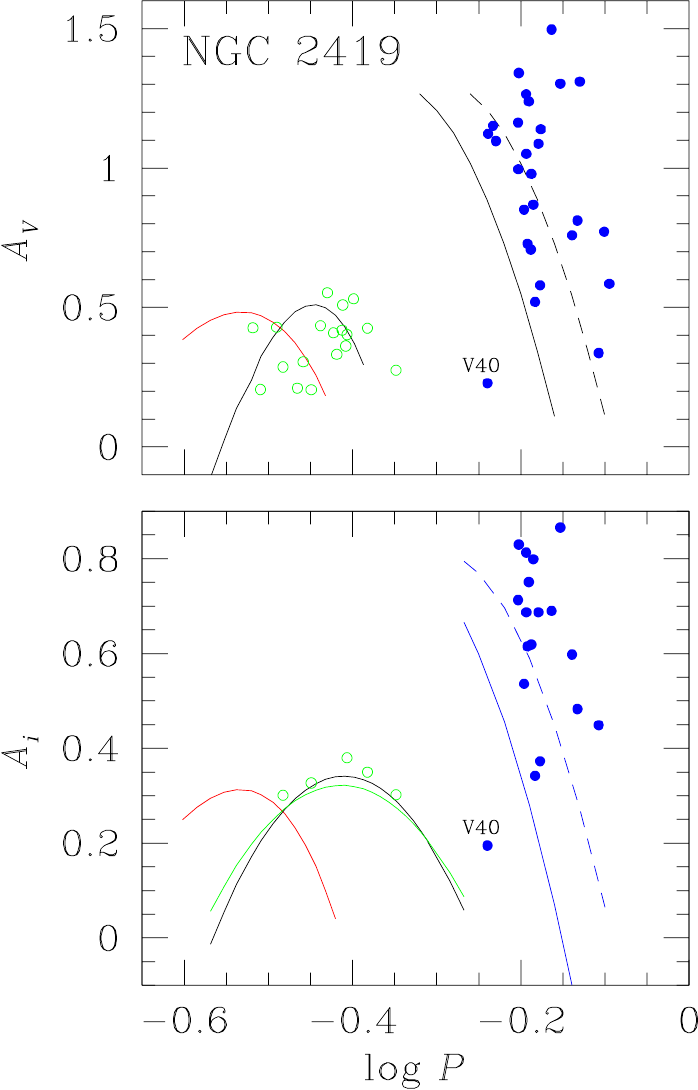}
\caption{ Period-Amplitude diagram for the RR Lyrae in NGC 2419. Blue and green symbols represent RRab and RRc stars respectively. The star V40 is not a cluster member.
 In the top panel, the curves to the right, are the locy for RRab stars (unevolved continuous and evolved segmented) in M3 according to \citep{Cacciari2005}. The black parabola for the RRc stars was calculated by \citet{Kunder2013b} from 14 OoII clusters and \citet{Arellano2015} calculated the red parabolas from a sample of RRc stars in five OoI clusters. In the bottom panel, the continuous and segmented blue lines were constructed by \citet{Kunder2013a}. The  black and green parabolas were calculated by \citet{Yepez20} and \citet{Deras2019} respectively, using 35 RRc stars from eight OoII clusters.}
\label{Bailey}
\end{center}
\end{figure}

Given its low metallicity the cluster likely  belongs to the Oo II type. This is in fact confirmed by the period average of its RRab and RRc stars of 0.665 d and 0.343 d respectively. The log P - Amplitude diagram (Bailey's diagram) is shown in Fig. \ref{Bailey}. It is clear that the distribution of stars, particularly that of the RRab, fall towards the sequences of evolved star \citep{Cacciari2005} which is the characteristic of Oo II type clusters.

\section{On the cluster reddening}

 Given its Galactic location, NGC 2419 is not much affected by interstellar reddening. To estimate the colour excess of individual RRab stars, we have taken advantage of the fact that the $V-I$ colour curve of RRab stars is constant near its minimum,  between the phases 0.5 and 0.8 \citep{Sturch1966}. By employing the calibration of $(V-I)_{0,min} = 0.58 \pm 0.02$ \citep{Guldenschuh2005}, we calculated the average colour excess for 12 cluster member RRab and found $E(B-V)=0.07 \pm0.07$. This average is comparable with the predicted  values from the calibrations of \citet{Schlafly2011} and \citet{Schlegel1998}, 
$0.052 \pm 0 .001$ and 	$0.061 \pm 0.001$ respectively. In this work we adopted $E(B-V)=0.06$ or $E(V-I)= 1.259 \times E(B-V)$, which leads to $E(V-I)= 0.08$.

\section{Physical Parameters of the RR Lyrae Stars from the light curves Fourier decomposition}
\label{sec:Four}

Our approach to the calculation of relevant stellar physical parameters for the RR Lyrae stars is via the Fourier decomposition of their $V$-band light curves.
Of particular interest are the 
mean [Fe/H] and $M_V$ of the RR Lyrae sample as they are good representations of the mean cluster metallicity and distance.

The Fourier decomposition is performed by fitting the
observed light curve with a Fourier series model of the form:

\begin{equation}
\label{eq.Foufit}
m(t) = A_0 + \sum_{k=1}^{N}{A_k \cos\ ({2\pi \over P}~k~(t-E) + \phi_k) },
\end{equation}

\noindent
where $m(t)$ is the magnitude at time $t$, $P$ is the period, and $E$ is the epoch. A
linear
minimization routine is used to derive the best-fit values of the 
amplitudes $A_k$ and phases $\phi_k$ of the sinusoidal components. 
From the amplitudes and phases of the harmonics in eq.~\ref{eq.Foufit}, the 
Fourier parameters, defined as $\phi_{ij} = j\phi_{i} - i\phi_{j}$, and $R_{ij} =
A_{i}/A_{j}$, are computed. 

Subsequently, the low-order Fourier parameters can be used in combination with
semi-empirical calibrations to calculate [Fe/H] and $M_V$. The employed calibrations for the mass, $T_{\rm eff}$, and radii are summarized in the papers by \citet{Arellano2010} and \citet{Arellano2011}, while the calibrations for [Fe/H] and $M_V$ and their zero points are most recently discussed by \citet{Arellano2022} and \citet{Arellano2024}.

\subsection{Physical parameters of RR Lyrae stars}
\begin{table*}[htp]
    \footnotesize
    \begin{center}
    \caption{Physical parameters from the member RR Lyrae Fourier light curve decomposition.}
    \begin{tabular}{ccccccccc} 
    \hline
    ID &[Fe/H]$_{\rm ZW}$ & [Fe/H]$_{\rm UVES}$  & $M_V$ & log~$T_{\rm eff}$  &log$(L/{L_{\odot}})$ &$M/{ M_{\odot}}$ & $D(kpc)$ &$R/{ R_{\odot}}$ \\
    \hline
    \multicolumn{9}{c}{RRab}\\
    \hline
    V3   &-1.78(9) & -1.77(11)   & 0.51(1) & 3.80(2) & 1.71(1) & 0.73(14) & 83.61(43) & 6.01(3) \\ 
    V9  &-1.74(9) & -1.77(11)   & 0.46(1) & 3.80(2) & 1.73(1) & 0.76(15) & 83.41(48) & 6.19(3) \\
    V12   &-2.09(10) & -2.24(15)   & 0.43(1) & 3.80(2) & 1.76(1) & 0.83(16) & 88.37(45) & 6.54(3) \\
     V13 &-1.69(10) & -1.65(12)   & 0.50(1) & 3.80(2) & 1.72(1) & 0.73(16) & 86.13(44) & 6.09(3) \\  
    V14  &-2.23(10)   & -2.47(17)  & 0.37(1) & 3.79(2) & 1.79(1) & 0.86(20) & 84.65(55) & 7.06(4) \\ 
    V15  &-1.71(11)  & -1.68(13)   & 0.46(1) & 3.80(2) & 1.73(1) & 0.76(17) & 88.41(52) & 6.18(3) \\   
    V16  &-2.05(18)     & -2.17(27)   & 0.34(2) & 3.80(3) & 1.79(1) & 0.89(31) & 94.75(1.12) & 6.75(7) \\    
    V17   &-1.91(12)  & -1.95(16)  & 0.45(1) & 3.80(2) & 1.74(1) & 0.78(19) & 87.28(62) & 6.30(4) \\  
    V19   &-1.88(12)    & -1.92(15)   & 0.35(1) & 3.80(2) & 1.78(1) & 0.80(19) & 92.56(55) & 6.65(4) \\    
    V21  &-1.91(10)   & -1.95(13)  & 0.39(1) & 3.80(2) & 1.77(1) & 0.80(19) & 86.50(51) & 6.58(3) \\   
    V22  &-1.79(10)  & -1.79(13)   & 0.47(1) & 3.81(3) & 1.73(1) & 0.80(26) & 93.55(68) & 5.94(4) \\   
    V24  &-1.57(15)  & -1.50(17)   & 0.49(1) & 3.80(3) & 1.72(1) & 0.68(24) & 86.77(63) & 5.99(4) \\  
    V25   &-1.74(7)    & -1.72(8)    & 0.51(1) & 3.80(1) & 1.72(1) & 0.72(12) & 82.27(29) & 6.03(2) \\    
    V29  &-2.12(12)   & -2.28(18)   & 0.40(01) & 3.79(1) & 1.77(1) & 0.84(18) & 87.01(46) & 6.90(3) \\
    V32   &-1.67(9)  & -1.62(10)    & 0.50(1) & 3.80(1) & 1.72(1) & 0.72(15) & 77.43(29) & 6.04(2) \\  
    V57  &-1.91(12)     & -1.97(16) & 0.46(1) & 3.79(2) & 1.74(1) & 0.74(20) & 75.11(33) & 6.64(3) \\    
    V59   &-2.51(62)$^1$   & -2.79(1.11)$^1$  & 0.39(02) & 3.78(6) & 1.79(1) & 0.82(60) & 89.19(88) & 7.27(6) \\
    \hline
    Mean &-1.86 & -1.90  & 0.44 & 3.80 & 1.75& 0.78 & 86.3 & 6.42\\
        $\sigma$ & $\pm 0.18$  &$\pm 0.27$ & $\pm 0.5$ & $\pm 0.01$ & $\pm 0.03$  & $\pm 0.03$ &  $\pm 5.0$ & $\pm 0.40$ \\
    \hline
    \multicolumn{8}{c}{RRc}\\
    \hline
    V4 &-1.71(11)  & -1.68(13)   & 0.51(4) & 3.83(1) & 1.69(1)  & 0.46(2)  & 85.43(14) & 4.54(7) \\
    V6 &-2.06(7)  & -2.19(11)   & 0.51(4) & 4.12(1) & 1.70(1)  & 0.53(2)  & 85.70(16) & 4.63(8) \\
    V27 &-1.54(14) & -1.45(16)   & 0.75(4) & 3.84(1) & 1.60(2)  & 0.53(2)  & 81.09(16) & 4.42(8) \\
    V31 &-1.61(20) & -1.55(23)   & 0.48(3) & 3.86(1) & 1.71(1) & 0.48(2)  & 85.60(14) & 4.59(8) \\
    V33 &-1.75(39) & -1.74(48)  & 0.55(8) & 3.84(1) & 1.68(3)  & 0.53(5)  & 92.40(35) & 4.91(18) \\    
    V34 &-1.70(8) & -1.66(10)    & 0.42(4) & 3.86(1) & 1.73(2)  & 0.50(3)  & 92.44(1.92) & 4.78(10) \\
    V55 &-1.06(20)$^1$ & -0.95(15)$^1$  & 0.52(5) & 3.87(1) & 1.69(2)  & 0.45(3)  & 85.79(1.67) & 4.39(10) \\   
    V60 &-1.60(3) & -1.53(3)  & 0.58(4) & 3.86(1) & 1.67(2)  & 0.42(2)  & 81.37(1.53) & 4.37(8) \\
    V66  & -1.79(11)  &-1.79(14) & 0.54(4) & 3.86(1) & 1.69(2)  & 0.46(2)  & 91.17(1.89) & 4.51(9) \\
    V67  & -1.96(5)  & -2.04(7)  & 0.56(3)&3.86(1) & 1.68(1)& 0.52(2)&70.60(1.03) & 4.46(6) \\
    V68 &-2.12(9)& -2.28(14)   & 0.48(2) & 3.85(1) & 1.71(1)  & 0.58(2) & 85.95(1.07) & 4.74(6) \\
    V72 &-2.12(10) & -2.29(15)   & 0.49(4) & 3.85(1) & 1.70(2)  & 0.51(2)  & 86.02(1.61) & 4.87(9) \\
    V74 &-2.05(8) & -2.17(12)   & 0.62(4) & 3.85(1) & 1.65(1)  & 0.61(3)  & 90.07(1.50) & 4.40(7) \\
    V75  &-1.36(6)$^1$& -1.25(6)$^1$   & 0.56(3) & 3.87(1) & 1.68(1)  & 0.53(2)  & 81.04(1.04) & 4.27(5) \\
    V76  &-1.49(11)& -1.39(12)  & 0.39(4) & 3.86(1) & 1.74(1)  & 0.45(2)  & 79.03(1.31) & 4.85(8) \\  
    V77  &-1.94(15)& -2.00(21)   & 0.44(4) & 3.84(1) & 1.72(1)  & 0.61(3)  & 86.16(1.43) & 5.09(8) \\  
    V84  &-2.05(15)& -2.19(22)   & 0.56(3) & 3.86(1) & 1.68(1)  & 0.58(2)  & 70.47(92)  & 4.48(6) \\
    V82 &-1.77(12) & -1.75(15)  & 0.77(3) & 3.84(1) & 1.59(1)  & 0.49(1)  & 60.06(75)  & 4.36(5) \\
    V90  &-2.085(24)& -2.23(35)  & 0.42(4) & 3.85(01) & 1.73(2)  & 0.56(3) & 88.57(1.83) & 4.88(10) \\
    \hline
    Mean & -1.84 &	-1.88 & 0.53 & 3.86 &  1.69 & 0.52& 83.1  & 4.61\\
    $\sigma$ & $\pm 0.21$  &$\pm 0.30$ & $\pm 0.10$ & $\pm 0.01$ & $\pm 0.04$  & $\pm 0.05$ &  $\pm 8.1$ & $\pm 0.23$ \\
    
    \hline
    \end{tabular}
    \center{$^1$ Value not included in the calculation of the mean.}\\
    \label{tab:parfisRR}
    \end{center}
\end{table*}

The Fourier coefficients for the RRab and RRc stars were calculated for the stars  that we were able to measure. For the sake of briefness these are not specifically given in this paper but they are available on request. These coefficients were then used in the cited calibrations to calculate the individual physical parameters, which are reported in Table \ref{tab:parfisRR}. The average [Fe/H] and $M_V$ (hence distance), should be representative of the parent cluster. It should be noted that the calibrations render the value [Fe/H]$_{\rm ZW}$, i.e. in the metallicity scale of \citet{Zinn1984}
which can be transformed into the spectroscopic scale of \citet{Carretta2009}, via the equation; [Fe/H]$_{\rm UVES}$= $-0.413$ + 0.130~[Fe/H]$_{\rm
ZW} - 0.356$~[Fe/H]$_{\rm ZW}^2$.

We found averages [Fe/H]$_{\rm ZW}= -1.86 \pm 0.18$, [Fe/H]$_{\rm UV}= -1.90 \pm 0.27$ and $D=86.3 \pm 5.0$ kpc. from the RRab light curves
and [Fe/H]$_{\rm ZW}= -1.84 \pm 0.21$, [Fe/H]$_{\rm UV}= -1.88 ~\pm~0.30$ and $D=83.1 \pm 8.1$ kpc from the RRc data. Emphasis should be made that,
coming from independent calibrations with different calibrators, the results for the RRab and RRc stars are also fully independent. Still
there is a very satisfactory agreement. These values can be compared with the value quoted [Fe/H]$_{Spec}=-2.2$ in the spectroscopic scale \citep{Carretta2009}, and the critical literature mean distance 
$D=88.47 \pm 2.40$ kpc estimated by \citet{Baumgardt2021}. Thus, the agreement of our metallicity and distance determinations for NGC 2419 with this canonical values, cannot be better.

\section{Discussion: The Color Magnitude Diagram}

\begin{figure*}[t]
\begin{center}
\includegraphics[width=16.0cm]{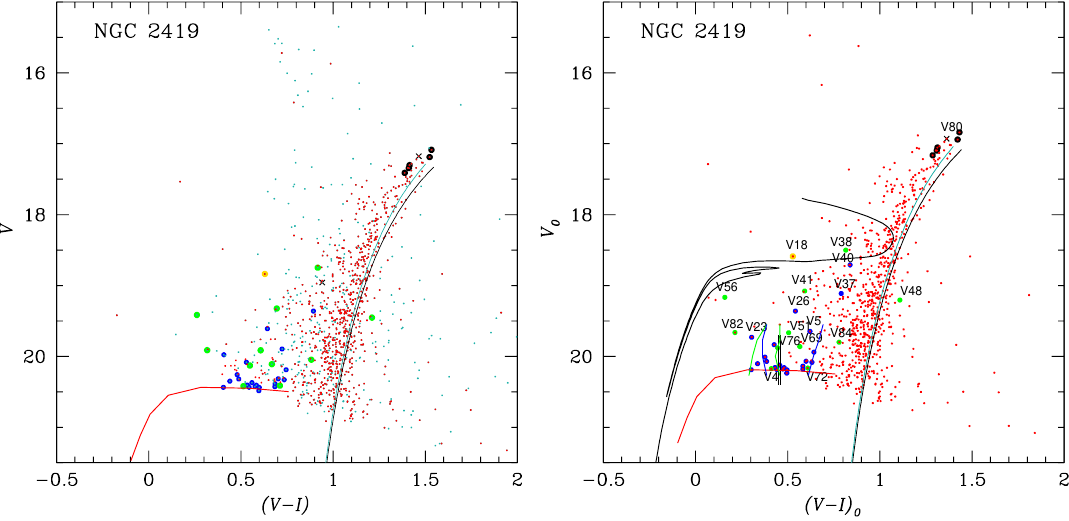}
\caption{Observed CMD of NGC 2419. Left panel shows the field stars (small blue dots) and likely cluster member stars (red dots), according to our membership analysis. The right panel shows only the member stars in the dereddened plane.
We use $E(B-V)=0.08$. Blue and green circles represent RRab and RRc stars, respectively. Black circles are used for SR stars. The only CW star is the yellow circle. Black X corresponds to V80, whose  variability we could not confirm. The vertical black lines at the ZAHB mark the empirical red edge of the first overtone instability strip  \citep{Arellano2015,Arellano2016}. Also shown are the theoretical fundamental (blue lines) and first overtone (green lines) instability strip borders from \citet{Bono1994}. Isochrones are from \citet{Vandenberg2014} for age of 12.5 Gyr (black) and 13.0 Gyr (turquoise) with [Fe/H]= -1.8 and -2.0, respectively. Red ZAHB is from the models built from the Eggleton code \citep{Pols1997,Pols1998, KPS1997}, and calculated by \citet{Yepez2022}. The black evolutionary track corresponds to a model of total mass 0.54 $M_{\odot}$ and a core of 0.50 $M_{\odot}$. These very thin envelope models explain well type II Cepheids such as V18 \citep{Yepez2022}. }
\label{CMD}
\end{center}
\end{figure*}

In Fig. \ref{CMD} we display two versions of the observed CMD. The panel on the left shows the stellar distribution of every point source that we were able to measure in our collection of \emph{VI} images. This includes cluster member stars as well as field stars. Through the membership analysis described in section \ref{sec:membership}, we could distinguish the most likely cluster members (red dots in the figure) from the field stars (small light blue dots).
In the right panel, we have plotted only the cluster members and have dereddened the diagram by assuming $E(B-V)=0.08$.
The variable stars contained in Table \ref{Vars} are plotted with colours according to their pulsational type; as coded in the figure caption. 
When a variable is found to be a cluster member, the colour symbol has a smaller red dot in the center, otherwise the star had no proper motion reported in $Gaia$-DR3 and hence its membership status is unknown, or it is likely a field star.

We also include in this dereddened diagram the 
two isochrones from \citet{Vandenberg2014} for ages of 12.5 Gyr (black) and 13.0 Gyr (turquoise) with [Fe/H]= -1.8 and -2.0, respectively. Red ZAHB is from the models built from the Eggleton code \citep{Pols1997,Pols1998, KPS1997}, and calculated by \citet{Yepez2022}. The black evolutionary track corresponds to a model with a central core mass of 0.50 $M_{\odot}$ and a thin envelope of 0.04 $M_{\odot}$ and it was selected as it best represents the position of the W Vir star V18 confirming the conclusions reached by \citet{Yepez2022}, that type II Cepheids can be interpreted as products of post-HB evolution driven by burning of very low mass hydrogen and helium shells. According to \citet{Bono2020}, W Vir stars may be post-early asymptotic giant branch stars (see also the case of the CWB star V81 in NGC 7006 \citep{Arellano2023}), although the case of V18 seems more like that of a post-HB which according to model has taken 114 million years from the HB to reach its present position.

A qualitative comparison of our limited CMD with the deep and detailed CMD, figure 2 of \citet{diCriscienzo2011} (hereinafter diC11), leads to the following
instructive  observations.
First, NGC 2419 has a well developed HB blue tail from where V18 has most likely evolved. 

Our photometry does not reach these faint regions of the CMD. If we go by the variable star distribution near the HB in diC11 CMD, we may conclude that all RR Lyrae in their study are cluster members. However, diC11 did not perform a membership analysis. In our CMD of Fig. \ref{CMD}, several RR Lyrae appear noticeably brighter than the HB (e.g. V37, V38, V40, V41, V48, V51, V56 and V69). It is not difficult to corroborate from the identification chart of Fig. \ref{IdChart}, that these stars are tenants of the cluster central regions and hence are most likely blended, making our photometry further limited and hence producing spurious positions in our CMD. These conditions may have also limited the possibilities of the $Gaia$ mission to measure them as none of them have proper motions reported in $Gaia$-DR3, thus, we have assigned them the unknown (UN) membership status in Table \ref{tab:datosgenerales1}. The membership of these variables in NGC 2419 should be corroborated in the future. The rest of the variables in Table 2, all likely members, are distributed closer to the HB with some scatter. This scatter is comparable to the observed in Figure 2 of diC11 if plotted at the same scale, therefore the RR Lyrae population shows signs of evolution off the ZAHB.

The mass distribution along the ZAHB is determined by the amount of mass lost during the He-flash events at RGB, as it is clearly demonstrated by the models of \citet{SilvaAguirre2008}. The more mass is lost, the less massive the star is when settling towards the bluer regions of the ZAHB. This is consistent for instance with the fact that bluer (hotter) RRc stars are less massive than redder (cooler) RRab stars. The two modes are in principle separated by the First Overtone Red Edge (FORE) of the instability strip. Depending upon exact total mass at exhaustion of core-Helium, RRc and RRab stars can either be neatly separated by the FORE or can share the inter-order or bimodal instability strip (i.e. the intersection of the first overtone and fundamental mode instability strips). These concepts are graphically illustrated in figure 3 of \citet{Caputo1978}.

The empirical position of the FORE  \citep{Arellano2015,Arellano2016} is shown in Fig. \ref{CMD} by two black vertical lines. Also shown are the theoretical borders of the instability strip for the first overtone and fundamental modes from \citet{Bono1994}. Note that the theoretical and empirical FORE match very well.
It was found by \citet{Arellano2019} that in all OoII type clusters studied by them, RR Lyrae pulsating modes are well segregated by the FORE, whereas such clear segregation happens only in some  OoI clusters while in others the two modes share the inter-order or "either-or" region. This was interpreted by these authors as in OoII clusters the RR Lyrae stars always start their ZAHB evolution from less massive bluer stars while in OoI clusters they exhibit a wider mass distribution and hence segregation may or may not occur. It remains unclear what physics constrain these two options but it is likely connected  with the mass-loss processes during the RGB.

In the case of NGC 2419, the RRc and RRab stars are all mixed across the HB. This can be seen in the CMD of Fig. \ref{CMD}, but also in the CMD in figure 2 of diC11. For a Oo II cluster, the location
  of RRab stars in the inter-order region is contrary to the argument
 given in the earlier paragraph. Hence, NGC 2419 is an unconventional Oo II cluster exhibiting mixing of the modes in the instability strip.

\section{Conclusions}

Using the $Gaia$-DR3 proper motions of stars in the field of NGC 2419, we have been able to confirm the membership of most of the variables known in the cluster. Several of the known variables are in crowded environments and due to blending with neighbours we were not able to  resolve them, and they may appear in odd positions in the CMD. These stars also lack of $Gaia$-DR3 proper motion data, thus we cannot conclude about their membership status
(e.g. V38, V41, V48, V51, V56, V69). V89, with proper motion measurement, was found to be a likely field star.

The radial distribution of member stars, clearly demonstrates the extended reach of the cluster to distances of about 140 pc, making of NGC 2419 one of the largest clusters in the Galaxy.

From the Fourier light curve decomposition of clearly resolved RR Lyrae stars, we determined the average metallicity and distance [Fe/H]$_{\rm UV}= -1.90 \pm 0.27$ and $D=86.3 \pm 5.0$ kpc from the RRab light curves
and [Fe/H]$_{\rm UV}= -1.88 \pm 0.30$ and $D=83.1 \pm 8.1$ kpc from the RRc stars. These determinations are in excellent agreement with the well established results [Fe/H]$_{Spec}=-2.2$ \citep{Carretta2009}, and  
$D=88.47 \pm 2.40$ kpc \citep{Baumgardt2021}.

Our post He-flash models show that the W Virginis star V18 has evolved from a ZAHB blue tail progenitor with a very thin shell; this progenitor has a total mass of 0.54 $M_{\odot}$
but a shell of only 0.04 $M_{\odot}$. These results confirm the conclusions from \citet{Deras2022} for the Pop II cepheids of M56 or from \citet{Yepez2022} for M14, that thin shells are a required condition for the generation of Pop II cepheids pulsations.

\vskip 1.0cm

\section{ACKNOWLEDGMENTS}

AAF is grateful to the Indian Institute of Astrophysics, for warm hospitality during the writing of this work.  AAF also thankfully acknowledges the sabbatical support granted by the program PASPA of the DGAPA-UNAM. The present project has been benefited from the support of DGAPA-UNAM through projects IG100620 and IN103024. The permanent help received from the IA-UNAM librarian, Beatriz Ju\'arez Santamar\'ia, with the bibliographical material needed for this work is fully acknowledged.   We are thankful to the IAO TACs for the telescope time allocations over 17 years and to the supporting staff in Hanle (IAO) and Hosakote (CREST) observing stations. The facilities at IAO and CREST are operated by the Indian Institute of Astrophysics, Bangalore.

\bibliography{NGC2419}

\begin{thebibliography}
\expandafter\ifx\csname natexlab\endcsname\relax\def\natexlab#1{#1}\fi
\expandafter\ifx\csname href\endcsname\relax
  \def\href#1#2{}\fi
\expandafter\ifx\csname urllinklabel\endcsname\relax
  \def\urllinklabel{[LINK]}\fi
\expandafter\ifx\csname adsurllinklabel\endcsname\relax
  \def\adsurllinklabel{[ADS]}\fi

\bibitem[{{Arellano Ferro}(2022)}]{Arellano2022}
{Arellano Ferro}, A. 2022, RevMexA\&A, 58, 257


\bibitem[{{Arellano Ferro}(2024)}]{Arellano2024}
---. 2024, IAU Symposium, 376, 222


\bibitem[{{Arellano Ferro} {et~al.}(2019){Arellano Ferro}, {Bustos Fierro}, {Calder{\'o}n}, \& {Ahumada}}]{Arellano2019}
{Arellano Ferro}, A., {Bustos Fierro}, I.~H., {Calder{\'o}n}, J.~H., \& {Ahumada}, J.~A. 2019, \rmxaa, 55, 337


\bibitem[{{Arellano Ferro} {et~al.}(2011){Arellano Ferro}, {Figuera Jaimes}, {Giridhar}, {Bramich}, {Hern{\'a}ndez Santisteban}, \& {Kuppuswamy}}]{Arellano2011}
{Arellano Ferro}, A., {Figuera Jaimes}, R., {Giridhar}, S., {Bramich}, D.~M., {Hern{\'a}ndez Santisteban}, J.~V., \& {Kuppuswamy}, K. 2011, \mnras, 416, 2265


\bibitem[{{Arellano Ferro} {et~al.}(2010){Arellano Ferro}, {Giridhar}, \& {Bramich}}]{Arellano2010}
{Arellano Ferro}, A., {Giridhar}, S., \& {Bramich}, D.~M. 2010, \mnras, 402, 226


\bibitem[{{Arellano Ferro} {et~al.}(2016){Arellano Ferro}, {Luna}, {Bramich}, {Giridhar}, {Ahumada}, \& {Muneer}}]{Arellano2016}
{Arellano Ferro}, A., {Luna}, A., {Bramich}, D.~M., {Giridhar}, S., {Ahumada}, J.~A., \& {Muneer}, S. 2016, \apss, 361, 175


\bibitem[{{Arellano Ferro} {et~al.}(2015){Arellano Ferro}, {Mancera Pi{\~n}a}, {Bramich}, {Giridhar}, {Ahumada}, {Kains}, \& {Kuppuswamy}}]{Arellano2015}
{Arellano Ferro}, A., {Mancera Pi{\~n}a}, P.~E., {Bramich}, D.~M., {Giridhar}, S., {Ahumada}, J.~A., {Kains}, N., \& {Kuppuswamy}, K. 2015, \mnras, 452, 727


\bibitem[{{Arellano Ferro} {et~al.}(2023){Arellano Ferro}, {Rojas Galindo}, {Bustos Fierro}, {Muneer}, {Yepez}, \& {Giridhar}}]{Arellano2023}
{Arellano Ferro}, A., {Rojas Galindo}, F.~C., {Bustos Fierro}, I.~H., {Muneer}, S., {Yepez}, M.~A., \& {Giridhar}, S. 2023, \mnras, 519, 2451


\bibitem[{{Baumgardt} \& {Vasiliev}(2021)}]{Baumgardt2021}
{Baumgardt}, H. \& {Vasiliev}, E. 2021, \mnras, 505, 5957


\bibitem[{{Bono} {et~al.}(2020){Bono}, {Braga}, {Fiorentino}, {Salaris}, {Pietrinferni}, {Castellani}, {Di Criscienzo}, {Fabrizio}, {Mart{\'\i}nez-V{\'a}zquez}, \& {Monelli}}]{Bono2020}
{Bono}, G., {Braga}, V.~F., {Fiorentino}, G., {Salaris}, M., {Pietrinferni}, A., {Castellani}, M., {Di Criscienzo}, M., {Fabrizio}, M., {Mart{\'\i}nez-V{\'a}zquez}, C.~E., \& {Monelli}, M. 2020, \aap, 644, A96


\bibitem[{{Bono} {et~al.}(1994){Bono}, {Caputo}, \& {Stellingwerf}}]{Bono1994}
{Bono}, G., {Caputo}, F., \& {Stellingwerf}, R.~F. 1994, \apj, 423, 294


\bibitem[{{Bramich}(2008)}]{Bramich2008}
{Bramich}, D.~M. 2008, \mnras, 386, L77


\bibitem[{{Bramich} {et~al.}(2015){Bramich}, {Bachelet}, {Alsubai}, {Mislis}, \& {Parley}}]{Bramich2015}
{Bramich}, D.~M., {Bachelet}, E., {Alsubai}, K.~A., {Mislis}, D., \& {Parley}, N. 2015, \aap, 577, A108


\bibitem[{{Bramich} {et~al.}(2011){Bramich}, {Figuera Jaimes}, {Giridhar}, \& {Arellano Ferro}}]{Bramich2011}
{Bramich}, D.~M., {Figuera Jaimes}, R., {Giridhar}, S., \& {Arellano Ferro}, A. 2011, MNRAS, 413, 1275


\bibitem[{{Bramich} {et~al.}(2013){Bramich}, {Horne}, {Albrow}, {Tsapras}, {Snodgrass}, {Street}, {Hundertmark}, {Kains}, {Arellano Ferro}, {Figuera}, \& {Giridhar}}]{Bramich2013}
{Bramich}, D.~M., {Horne}, K., {Albrow}, M.~D., {Tsapras}, Y., {Snodgrass}, C., {Street}, R.~A., {Hundertmark}, M., {Kains}, N., {Arellano Ferro}, A., {Figuera}, J.~R., \& {Giridhar}, S. 2013, \mnras, 428, 2275


\bibitem[{{Bustos Fierro} \& {Calder{\'o}n}(2019)}]{Bustos2019}
{Bustos Fierro}, I.~H. \& {Calder{\'o}n}, J.~H. 2019, \mnras, 488, 3024


\bibitem[{{Cacciari} {et~al.}(2005){Cacciari}, {Corwin}, \& {Carney}}]{Cacciari2005}
{Cacciari}, C., {Corwin}, T.~M., \& {Carney}, B.~W. 2005, \aj, 129, 267


\bibitem[{{Caputo} {et~al.}(1978){Caputo}, {Castellani}, \& {Tornambe}}]{Caputo1978}
{Caputo}, F., {Castellani}, V., \& {Tornambe}, A. 1978, \aap, 67, 107


\bibitem[{{Carretta} {et~al.}(2009){Carretta}, {Bragaglia}, {Gratton}, {D'Orazi}, \& {Lucatello}}]{Carretta2009}
{Carretta}, E., {Bragaglia}, A., {Gratton}, R., {D'Orazi}, V., \& {Lucatello}, S. 2009, \aap, 508, 695


\bibitem[{{Catelan}(2009)}]{Catelan2009}
{Catelan}, M. 2009, \apss, 320, 261


\bibitem[{{Clement} \& {Nemec}(1990)}]{Clement1990}
{Clement}, C. \& {Nemec}, J. 1990, \jrasc, 84, 434


\bibitem[{{Clement} {et~al.}(2001){Clement}, {Muzzin}, {Dufton}, {Ponnampalam}, {Wang}, {Burford}, {Richardson}, {Rosebery}, {Rowe}, \& {Hogg}}]{Clement2001}
{Clement}, C.~M., {Muzzin}, A., {Dufton}, Q., {Ponnampalam}, T., {Wang}, J., {Burford}, J., {Richardson}, A., {Rosebery}, T., {Rowe}, J., \& {Hogg}, H.~S. 2001, \aj, 122, 2587


\bibitem[{{Deras} {et~al.}(2022){Deras}, {Arellano Ferro}, {Bustos Fierro}, \& {Yepez}}]{Deras2022}
{Deras}, D., {Arellano Ferro}, A., {Bustos Fierro}, I., \& {Yepez}, M.~A. 2022, \rmxaa, 58, 121


\bibitem[{{Deras} {et~al.}(2019){Deras}, {Arellano Ferro}, {L{\'a}zaro}, {Bustos Fierro}, {Calder{\'o}n}, {Muneer}, \& {Giridhar}}]{Deras2019}
{Deras}, D., {Arellano Ferro}, A., {L{\'a}zaro}, C., {Bustos Fierro}, I.~H., {Calder{\'o}n}, J.~H., {Muneer}, S., \& {Giridhar}, S. 2019, \mnras, 486, 2791


\bibitem[{{Di Criscienzo} {et~al.}(2011){Di Criscienzo}, {Greco}, {Ripepi}, {Clementini}, {Dall'Ora}, {Marconi}, {Musella}, {Federici}, \& {Di Fabrizio}}]{diCriscienzo2011}
{Di Criscienzo}, M., {Greco}, C., {Ripepi}, V., {Clementini}, G., {Dall'Ora}, M., {Marconi}, M., {Musella}, I., {Federici}, L., \& {Di Fabrizio}, L. 2011, \aj, 141, 81


\bibitem[{{Gaia Collaboration} {et~al.}(2023){Gaia Collaboration}, {Vallenari}, {Brown}, {Prusti}, \& et~al.}]{Gaia2023}
{Gaia Collaboration}, {Vallenari}, A., {Brown}, A.~G.~A., {Prusti}, T., \& et~al. 2023, \aap, 674, A1


\bibitem[{{Guldenschuh} {et~al.}(2005){Guldenschuh}, {Layden}, {Wan}, {Whiting}, {van der Bliek}, {Baca}, {Carlin}, {Freismuth}, {Mora}, {Salyk}, {Vera}, {Verdugo}, \& {Young}}]{Guldenschuh2005}
{Guldenschuh}, K.~A., {Layden}, A.~C., {Wan}, Y., {Whiting}, A., {van der Bliek}, N., {Baca}, P., {Carlin}, J., {Freismuth}, T., {Mora}, M., {Salyk}, C., {Vera}, S., {Verdugo}, M., \& {Young}, A. 2005, \pasp, 117, 721


\bibitem[{{Kunder} {et~al.}(2013{\natexlab{a}}){Kunder}, {Stetson}, {Cassisi}, {Layden}, {Bono}, {Catelan}, {Walker}, {Paredes Alvarez}, {Clem}, {Matsunaga}, {Salaris}, {Lee}, \& {Chaboyer}}]{Kunder2013a}
{Kunder}, A., {Stetson}, P.~B., {Cassisi}, S., {Layden}, A., {Bono}, G., {Catelan}, M., {Walker}, A.~R., {Paredes Alvarez}, L., {Clem}, J.~L., {Matsunaga}, N., {Salaris}, M., {Lee}, J.-W., \& {Chaboyer}, B. 2013{\natexlab{a}}, \aj, 146, 119


\bibitem[{{Kunder} {et~al.}(2013{\natexlab{b}}){Kunder}, {Stetson}, {Catelan}, {Walker}, \& {Amigo}}]{Kunder2013b}
{Kunder}, A., {Stetson}, P.~B., {Catelan}, M., {Walker}, A.~R., \& {Amigo}, P. 2013{\natexlab{b}}, \aj, 145, 33


\bibitem[{{Landolt}(1992)}]{Landolt1992}
{Landolt}, A.~U. 1992, \aj, 104, 340


\bibitem[{{Pols} {et~al.}(1998){Pols}, {Schr{\"o}der}, {Hurley}, {Tout}, \& {Eggleton}}]{Pols1998}
{Pols}, O.~R., {Schr{\"o}der}, K.-P., {Hurley}, J.~R., {Tout}, C.~A., \& {Eggleton}, P.~P. 1998, MNRAS, 298, 525


\bibitem[{{Pols} {et~al.}(1997){Pols}, {Tout}, {Schroder}, {Eggleton}, \& {Manners}}]{Pols1997}
{Pols}, O.~R., {Tout}, C.~A., {Schroder}, K.-P., {Eggleton}, P.~P., \& {Manners}, J. 1997, MNRAS, 289, 869


\bibitem[{{Ripepi} {et~al.}(2007){Ripepi}, {Clementini}, {Di Criscienzo}, {Greco}, {Dall'Ora}, {Federici}, {Di Fabrizio}, {Musella}, {Marconi}, {Baldacci}, \& {Maio}}]{Ripepi2007}
{Ripepi}, V., {Clementini}, G., {Di Criscienzo}, M., {Greco}, C., {Dall'Ora}, M., {Federici}, L., {Di Fabrizio}, L., {Musella}, I., {Marconi}, M., {Baldacci}, L., \& {Maio}, M. 2007, \apjl, 667, L61


\bibitem[{{Schlafly} \& {Finkbeiner}(2011)}]{Schlafly2011}
{Schlafly}, E.~F. \& {Finkbeiner}, D.~P. 2011, \apj, 737, 103


\bibitem[{{Schlegel} {et~al.}(1998){Schlegel}, {Finkbeiner}, \& {Davis}}]{Schlegel1998}
{Schlegel}, D.~J., {Finkbeiner}, D.~P., \& {Davis}, M. 1998, \apj, 500, 525


\bibitem[{{Schr{\"o}der} {et~al.}(1997){Schr{\"o}der}, {Pols}, \& {Eggleton}}]{KPS1997}
{Schr{\"o}der}, K.-P., {Pols}, O.~R., \& {Eggleton}, P.~P. 1997, MNRAS, 285, 696


\bibitem[{{Silva Aguirre} {et~al.}(2008){Silva Aguirre}, {Catelan}, {Weiss}, \& {Valcarce}}]{SilvaAguirre2008}
{Silva Aguirre}, V., {Catelan}, M., {Weiss}, A., \& {Valcarce}, A.~A.~R. 2008, \aap, 489, 1201


\bibitem[{{Stetson}(2000)}]{Stetson2000}
{Stetson}, P.~B. 2000, \pasp, 112, 925


\bibitem[{{Sturch}(1966)}]{Sturch1966}
{Sturch}, C. 1966, \apj, 143, 774


\bibitem[{{van den Bergh} \& {Mackey}(2004)}]{vdB2004}
{van den Bergh}, S. \& {Mackey}, A.~D. 2004, \mnras, 354, 713


\bibitem[{{VandenBerg} {et~al.}(2014){VandenBerg}, {Bergbusch}, {Ferguson}, \& {Edvardsson}}]{Vandenberg2014}
{VandenBerg}, D.~A., {Bergbusch}, P.~A., {Ferguson}, J.~W., \& {Edvardsson}, B. 2014, \apj, 794, 72


\bibitem[{{Yepez} {et~al.}(2020){Yepez}, {Arellano Ferro}, \& {Deras}}]{Yepez20}
{Yepez}, M.~A., {Arellano Ferro}, A., \& {Deras}, D. 2020, \mnras, 494, 3212


\bibitem[{{Yepez} {et~al.}(2022){Yepez}, {Arellano Ferro}, {Deras}, {Bustos Fierro}, {Muneer}, \& {Schr{\"o}der}}]{Yepez2022}
{Yepez}, M.~A., {Arellano Ferro}, A., {Deras}, D., {Bustos Fierro}, I., {Muneer}, S., \& {Schr{\"o}der}, K.~P. 2022, \mnras


\bibitem[{Zhang {et~al.}(1996)Zhang, Ramakrishnan, \& Livny}]{Zhang1996}
Zhang, T., Ramakrishnan, R., \& Livny, M. 1996, SIGMOD Rec., 25, 103
 \href{http://doi.acm.org/10.1145/235968.233324}{\urllinklabel}

\bibitem[{{Zinn} \& {West}(1984)}]{Zinn1984}
{Zinn}, R. \& {West}, M.~J. 1984, \apjs, 55, 45


\end{thebibliography}

\appendix
\section{APPENDIX}
\setcounter{figure}{0}
\counterwithin{figure}{section}

\subsection{Light curves of measured variable stars}

The light curves of all variables resolved in our photometry are displayed in Figs. \ref{Mosai_RRab}, \ref{Mosai_RRc}, \ref{Mosai_SR} and \ref{V18} for the RRab, RRc, RGBs and CW respectively.

\begin{figure*}[ht]
\begin{center}
\includegraphics[width=16.0cm]{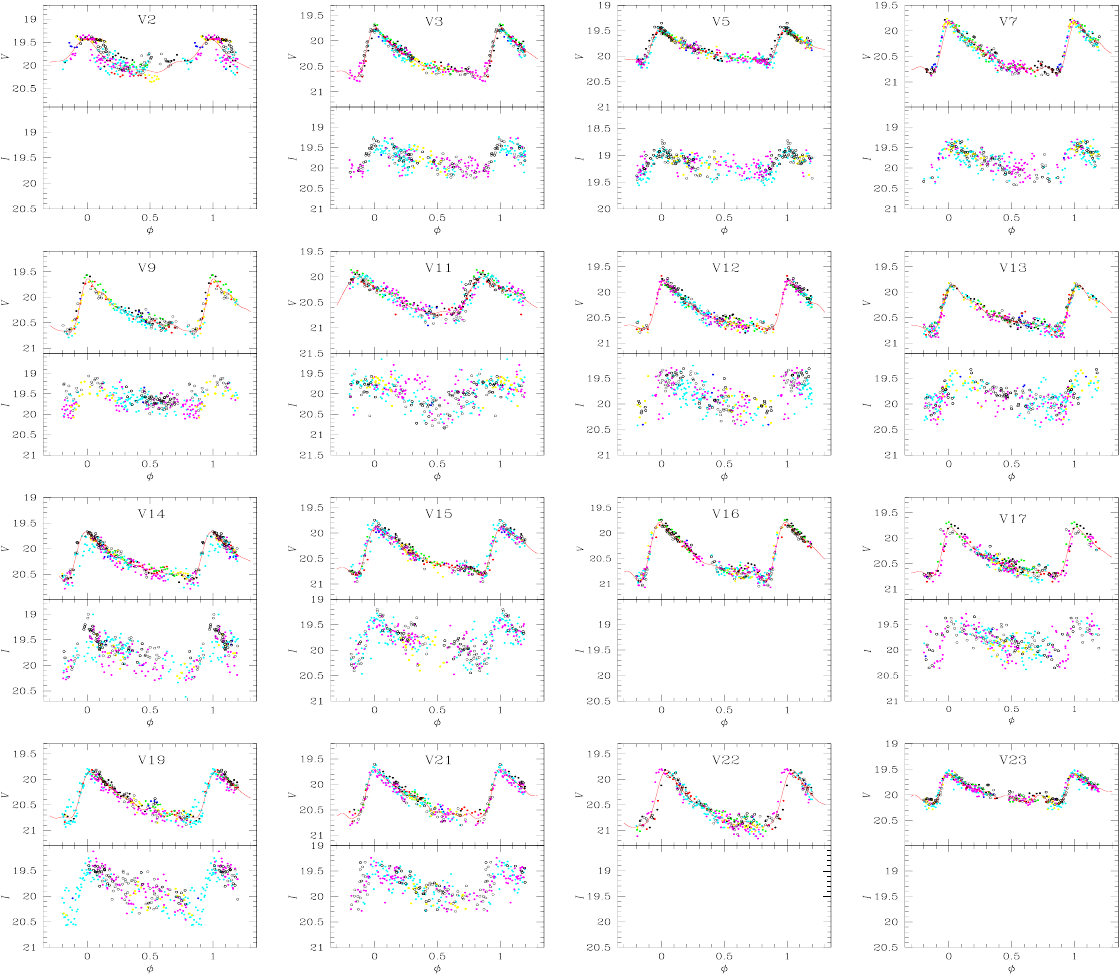}
\caption{Light curves of the RRab stars. Color code is as follows: black, red, green, blue, turquoise, lilac, yellow and empty circles for 2005, March 2007, April 2007, 2009, 2011-2012, 2013, 2017 and 2021-2022 seasons, respectively.}
\label{Mosai_RRab}
\end{center}
\end{figure*}

\begin{figure*}[htp]
\begin{center}
\includegraphics[width=16.0cm]{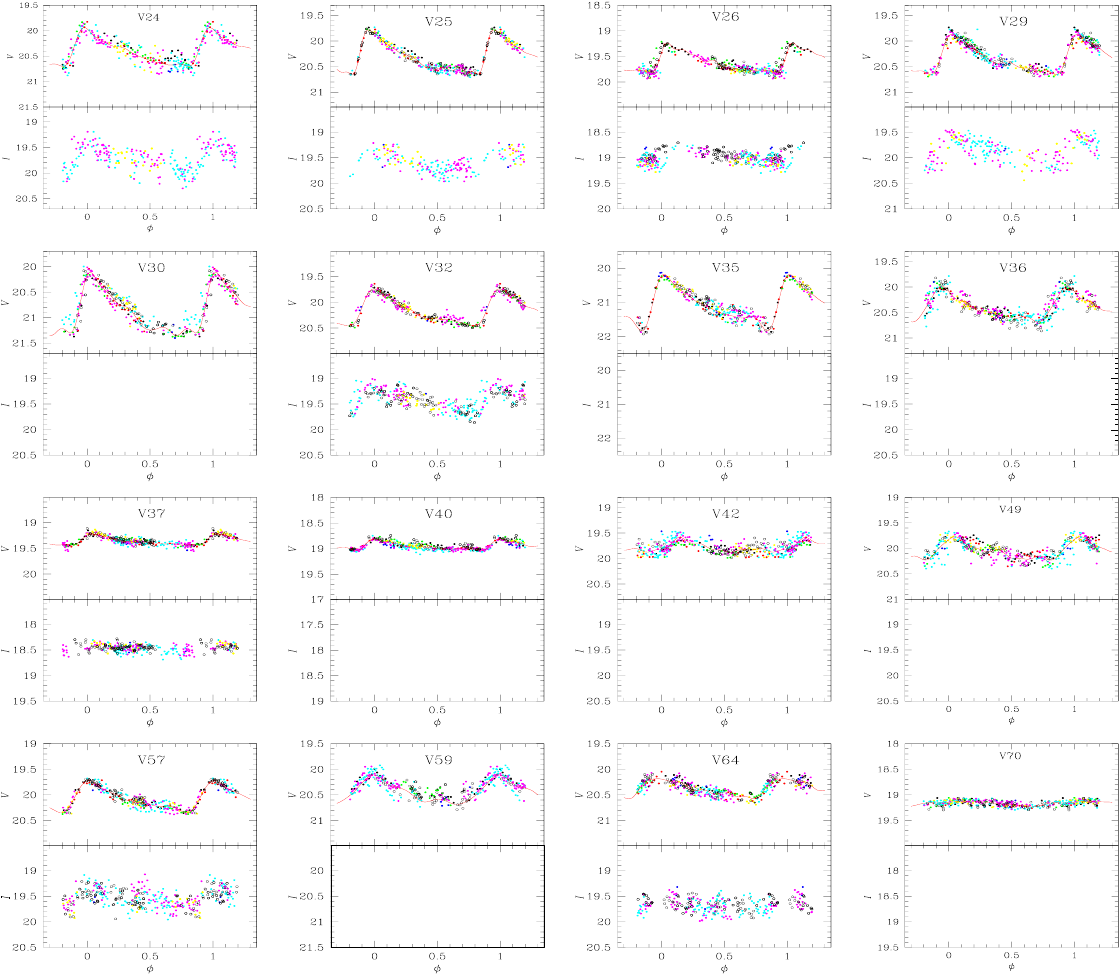}
\addtocounter{figure}{-1}
\caption{\small Continue}
\end{center}
\end{figure*}

\begin{figure*}[htp]
\begin{center}
\includegraphics[width=16.0cm]{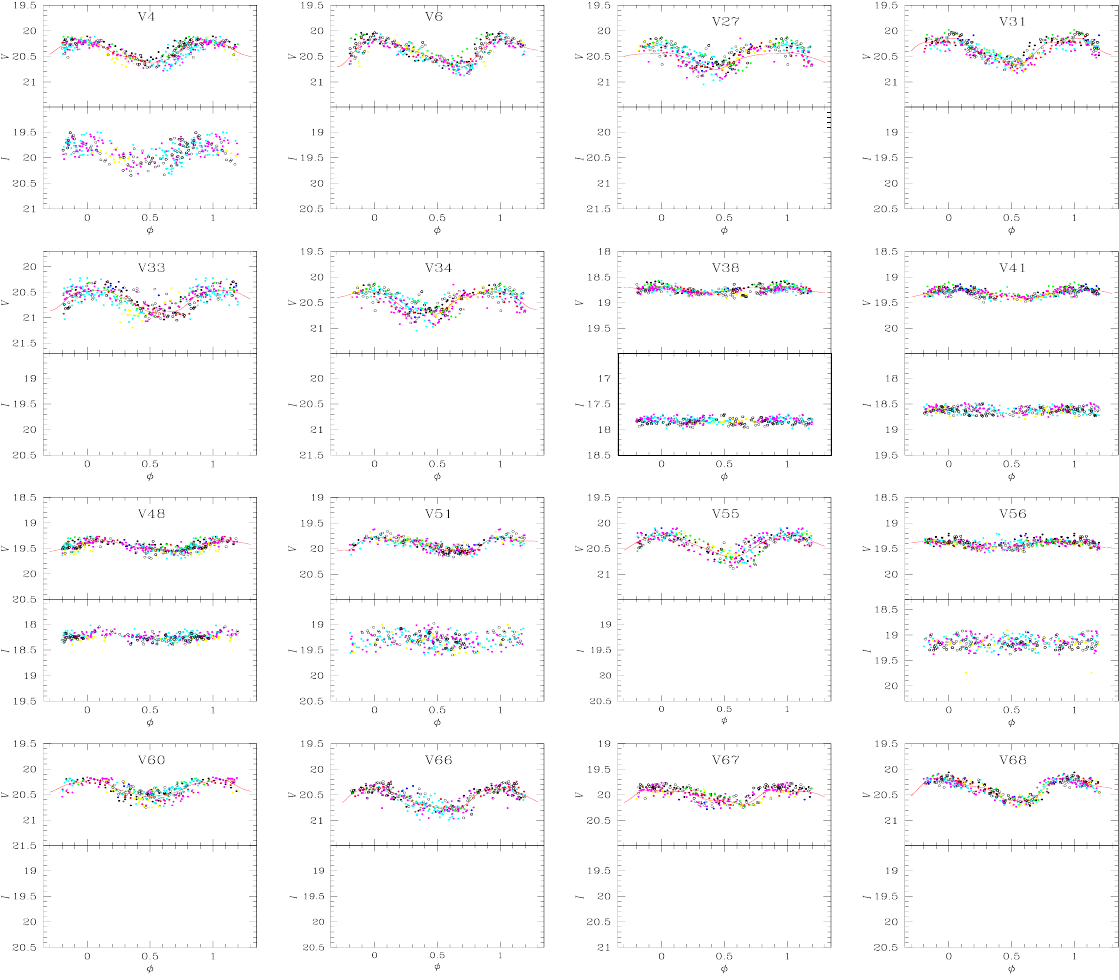}
\caption{Light curves of the RRc stars. Color code is as Fig. \ref{Mosai_RRab}}
\label{Mosai_RRc}
\end{center}
\end{figure*}

\begin{figure*}[htp]
\begin{center}
\includegraphics[width=16.0cm]{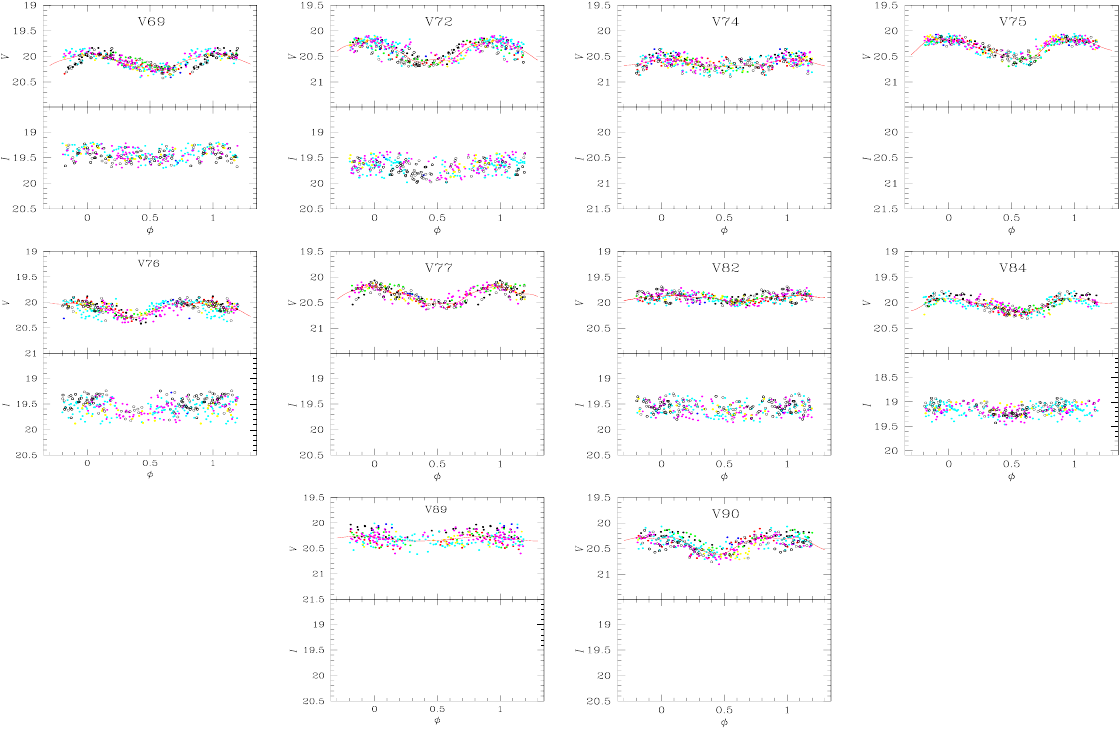}
\addtocounter{figure}{-1}
\caption{\small Continue}
\end{center}
\end{figure*}

\begin{figure*}[ht]
\begin{center}
\includegraphics[width=8.0cm]{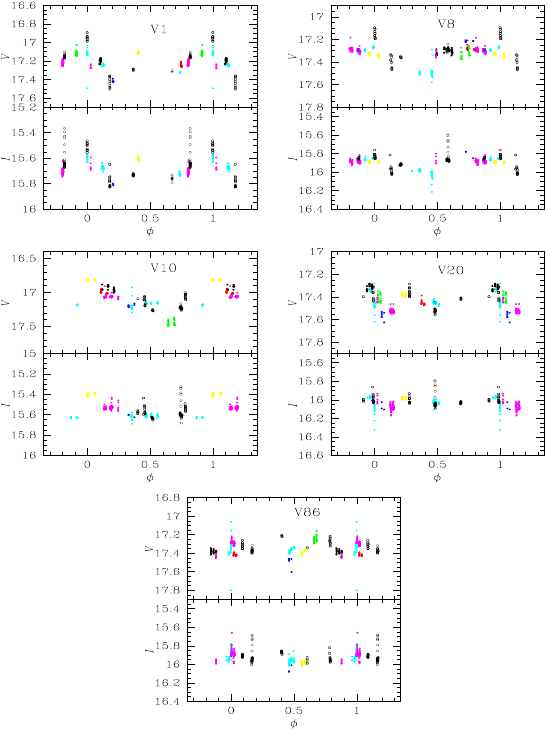}
\caption{Light curves of Long period variables in NGC 2419 phased with the periods of Table \ref{tab:datosgenerales1}. Color code is as Fig. \ref{Mosai_RRab}}
\label{Mosai_SR}
\end{center}
\end{figure*}

\subsection{Comments on individual stars}

V37, V49 and V42. These three stars display a very low amplitude in V.
This has also been noted by diC11 and \citet{Clement2001}. These stars are in the central regions of the cluster and then are very likely blended with near neighbours, so that their amplitudes appear diminished. None of them have proper motions reported in $Gaia$-DR3, thus we cannot confirm their cluster membership.

V38, V41, V48, V51, V56, V69. These RRc variables are all blended in our images. They have proper motions reported in $Gaia$-DR3, thus no membership status could be assigned.

V39. This double mode star was first found by \citet{Clement1990} with the periods $P_1=0.40704$ and $P_0=0.5465$. The fitting of our data with a model of these two periods is shown in Fig. \ref{V39} and looks quite satisfactory given the intrinsic noise of our observations.  

\begin{figure}
\begin{center}
\includegraphics[width=7.8cm]{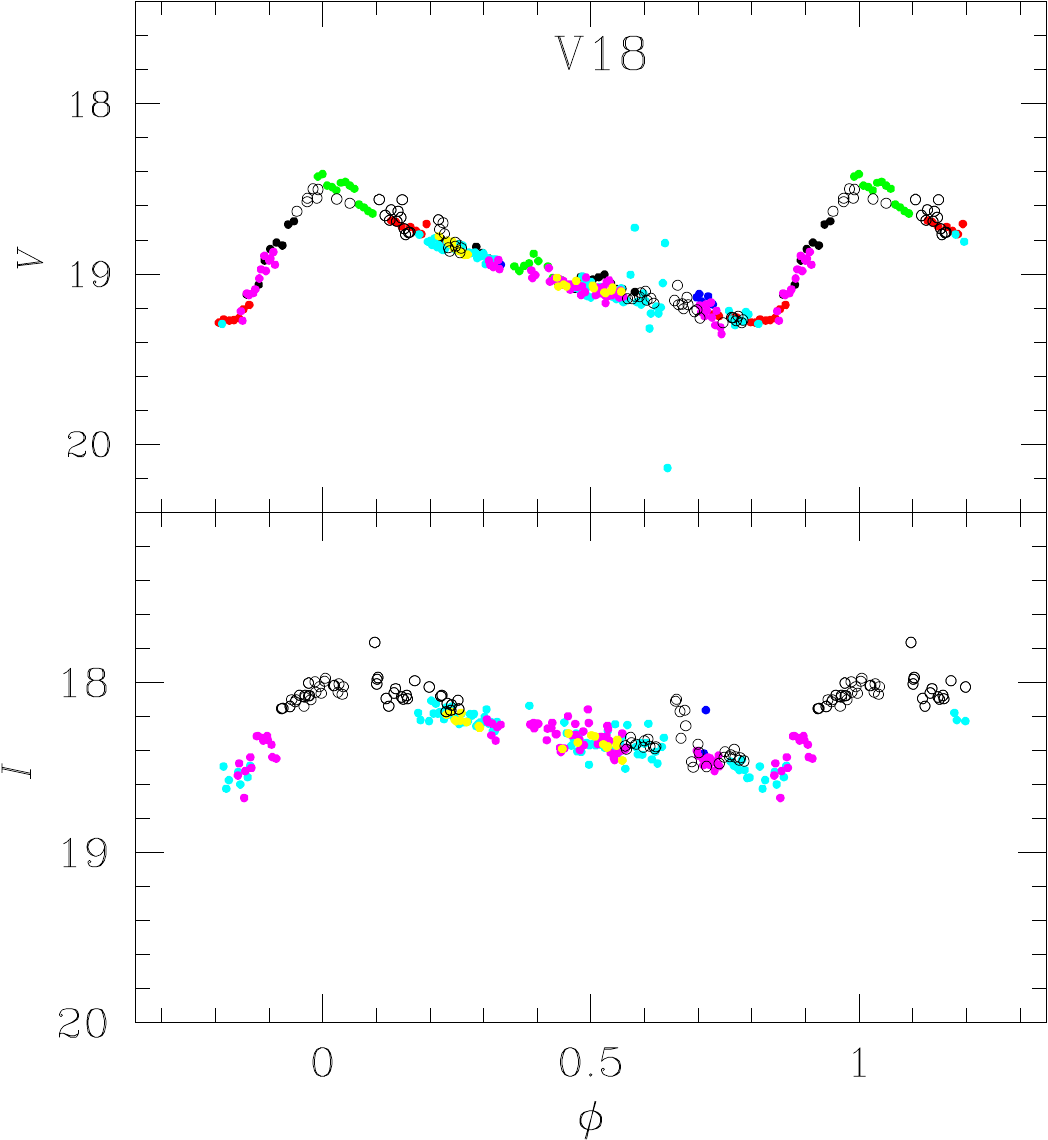}
\caption{Light curve of the CW star V18}
\label{V18}
\end{center}
\end{figure}

\begin{figure}
\begin{center}
\includegraphics[width=7.8cm]{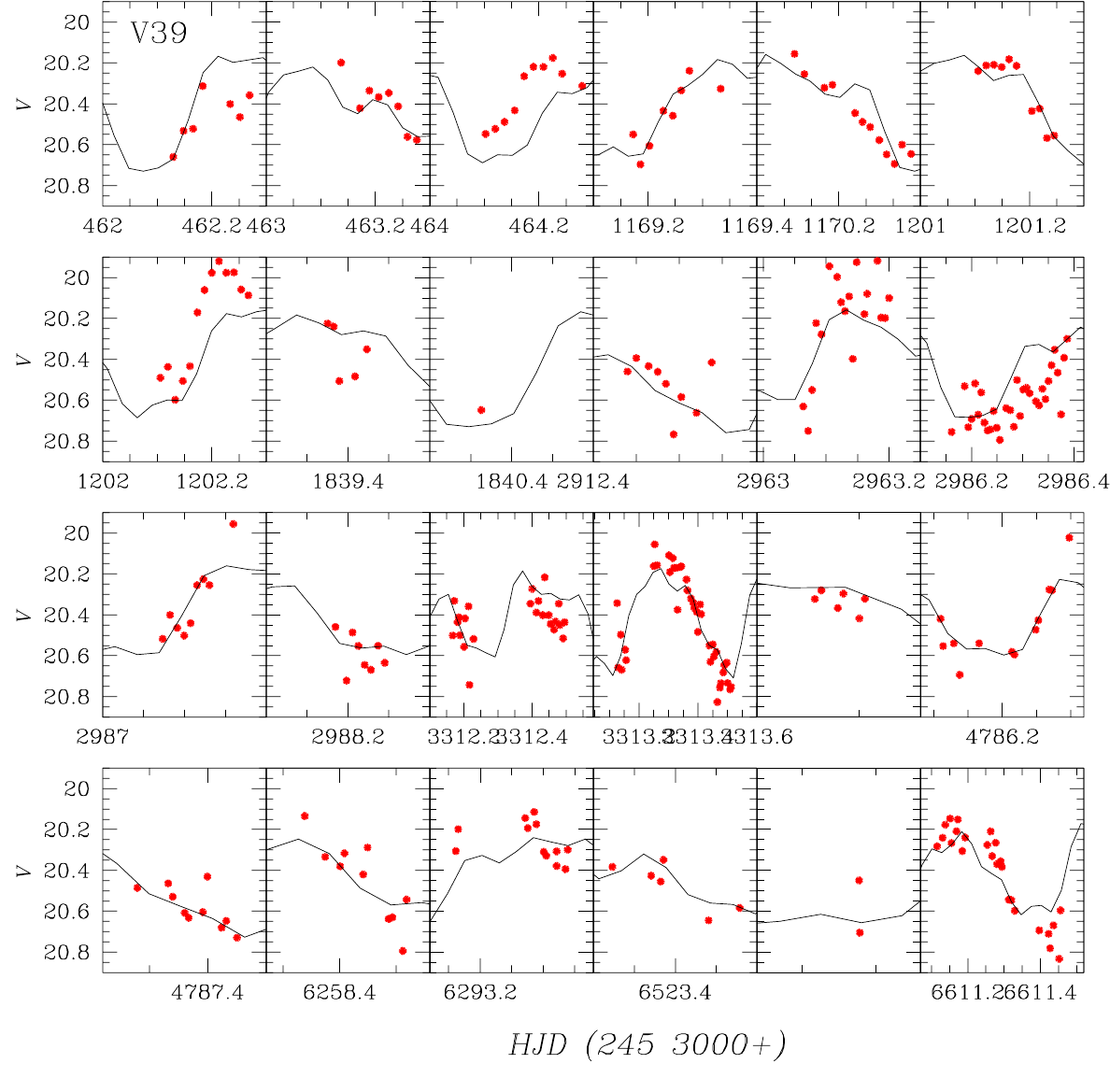}
\caption{Light curve of V39 with a two-period model fit with $P_0=0.54650$ and $P_1=0.40704$.}
\label{V39}
\end{center}
\end{figure}

\end{document}